\documentclass[12pt,journal,onecolumn]{IEEEtran}
\IEEEoverridecommandlockouts

\usepackage{epsfig,rotating,setspace,latexsym,amsmath,epsf,amssymb,amsfonts,bm,theorem,cite,algorithm,graphicx,epsf,authblk,epstopdf,color,algpseudocode,bbm,subcaption}

\newtheorem{remark}{Remark}

\newtheorem{theorem}{Theorem}
\newtheorem{lemma}{Lemma}

\allowdisplaybreaks

\begin{document}

\title{Sample, Quantize and Encode:\\Timely Estimation Over Noisy Channels\thanks{This work was supported in part by the U.S. National Science Foundation under Grant CCF-1908308. This article has been presented in part at the 2020 International Symposium of Information Theory (ISIT), Los Angeles, CA, June 2020 \cite{arafa-aoi-estimate-ou}.}\thanks{Ahmed Arafa is with the Department of Electrical and Computer Engineering, University of North Carolina at Charlotte, USA. Email: {\it aarafa@uncc.edu}.}\thanks{Karim Banawan is with the Department of Electrical Engineering, Alexandria University, Egypt. Email: {\it kbanawan@alexu.edu.eg}.}\thanks{Karim G. Seddik is with the Electronics and Communications Engineering Department, American University in Cairo, Egypt. Email: {\it kseddik@aucegypt.edu}.}\thanks{H. Vincent Poor is with the Electrical and Computer Engineering Department, Princeton University, USA. Email: {\it poor@princeton.edu}.}}

\author{Ahmed Arafa,~\IEEEmembership{Member,~IEEE}, Karim Banawan,~\IEEEmembership{Member,~IEEE}, Karim G. Seddik,~\IEEEmembership{Senior Member,~IEEE}, and H. Vincent Poor,~\IEEEmembership{Fellow,~IEEE}}

\maketitle

\begin{abstract}
The effects of {\it quantization} and {\it coding} on the estimation quality of Gauss-Markov processes are considered, with a special attention to the Ornstein-Uhlenbeck process. Samples are acquired from the process, quantized, and then encoded for transmission using either {\it infinite incremental redundancy} (IIR) or {\it fixed redundancy} (FR) coding schemes. A fixed {\it processing} time is consumed at the receiver for decoding and sending feedback to the transmitter. Decoded messages are used to construct a minimum mean square error (MMSE) estimate of the process as a function of time. This is shown to be an increasing functional of the {\it age-of-information} (AoI), defined as the time elapsed since the sampling time pertaining to the latest successfully decoded message. Such functional depends on the quantization bits, codewords lengths and receiver processing time. The goal, for each coding scheme, is to optimize sampling times such that the long-term average MMSE is minimized. This is then characterized in the setting of {\it general increasing functionals of AoI,} not necessarily corresponding to MMSE, which may be of independent interest in other contexts. 

We first show that the optimal sampling policy for IIR is such that a new sample is generated only if the AoI exceeds a certain {\it threshold,} while for FR it is such that a new sample is delivered {\it just-in-time} as the receiver finishes processing the previous one. {\it Enhanced} transmissions schemes are then developed in order to exploit the processing times to make new data available at the receiver sooner. For both IIR and FR, it is shown that there exists an optimal number of quantization bits that balances AoI and quantization errors, and hence minimizes the MMSE. It is also shown that for longer receiver processing times, the relatively simpler FR scheme outperforms IIR.
\end{abstract}


\section{Introduction}

Recent works have drawn connections between remote estimation of a time-varying process and the {\it age-of-information} (AoI) metric, which assesses the timeliness and freshness of the estimated data. While most works focus on transmitting {\it analog} samples for the purpose of estimation, this work focuses on using {\it quantized} and {\it coded} samples in that regard. We present optimal sampling methods that minimize the long-term average minimum mean square error (MMSE) of a Gauss-Markov, namely Ornstein-Uhlenbeck (OU), process under specific coding schemes, taking into consideration receiver {\it processing} times consumed in decoding and sending feedback. The OU process is the continuous-time analogue of the first-order autoregressive process \cite{ou-brownian-motion, doob-brownian-motion}, and is used to model various physical phenomena, and has relevant applications in control and finance. Our goal in this work is to devise practical sampling and coding schemes for the purpose of real-time tracking of OU processes while taking into consideration the effects of quantization, coding delays, and receiver processing times.

AoI, or merely age, is a time-based metric that measures information freshness by capturing delay from the receiver's perspective; it is defined as the time elapsed since the latest received data at the destination has been generated at its source. Hence, in general, to keep the data fresh, one needs to keep the AoI low. An increasing number of works in the recent literature have used AoI as a latency performance metric in various contexts. These include queuing-theoretic analyses of AoI for single and multiple sources \cite{yates_age_1, ephremides_age_random, ephremides_age_management, ephremides_age_non_linear, yates-age-mltpl-src, talak-aoi-delay, inoue-aoi-general-formula-fcfs, soysal-aoi-gg11, zou-waiting-aoi}, scheduling and sampling for AoI minimization \cite{modiano-age-bc, sun-age-mdp, zhou-age-iot, sun-cyr-aoi-non-linear, tang-aoi-power-multi-state}, status updating under energy harvesting constraints \cite{yates_age_eh, jing-age-online, baknina-updt-info, arafa-age-online-finite, bacinoglu-aoi-eh-finite-gnrl-pnlty, leng-aoi-eh-cog-radio}, AoI analysis in multihop networks \cite{batu-aoi-multihop, bedewy-aoi-multihop}, source coding for AoI minimization \cite{himanshu-age-source-coding}, and using AoI in other applications such as fresh data pricing \cite{zhang-arafa-aoi-pricing-wiopt}, cloud computing \cite{arafa-aoi-compute} and federated learning \cite{yang-arafa-aoi-fl}, among others, see the recent survey in \cite{aoi-survey-jsac}.

There are two main lines of research in the AoI literature that relate to this work. The first is the one pertaining to coding over noisy channels for age minimization, e.g., \cite{najm-age-mg11-harq, sac-age-mg1-harq, simeone-age-finite-code, yates-age-erase-code, baknina-age-coding, feng-age-rateless-codes, najm-age-erasure-coding, javani-aoi-erasure, parag-age-coding, ceran-age-harq, arafa-aoi-coding, huang-estimation-harq-control, chen-aoi-coding-bc, feng-coding-aoi-bc}. These works can be categorized according to the structure of the code being used to transmit the samples; references \cite{najm-age-mg11-harq, sac-age-mg1-harq, simeone-age-finite-code, yates-age-erase-code, baknina-age-coding, feng-age-rateless-codes, najm-age-erasure-coding, javani-aoi-erasure} focus on analyzing the usage of (rateless) infinite incremental redundancy (IIR) and fixed redundancy (FR) coding schemes and determined conditions in which both perform relatively well; the works in \cite{parag-age-coding, ceran-age-harq, arafa-aoi-coding, huang-estimation-harq-control} analyze the usage of hybrid ARQ (HARQ) coding schemes for AoI minimization; while those in \cite{chen-aoi-coding-bc, feng-coding-aoi-bc} consider broadcast multi-user settings. In IIR schemes, the transmitter sends its messages using a codewords of some original length, and then adds incremental redundancy (IR) bits one by one when signaled by the receiver until successful decoding is accomplished. This may potentially take a very large number of IR bit transmissions, hence the name IIR.\footnote{A clear example of the IIR scheme is the family of fountain (rateless) codes. In a rateless code, the encoder produces limitless (potentially infinite) stream of coded symbols based on the $\ell$ input symbols. The decoder reconstructs the $\ell$ bits after receiving \emph{any} $n$ correct symbols. One common rateless code is the systematic Raptor code in \cite{raptorRFC}, which is used in the 3GPP multimedia broadcast multicast services (MBMS), DVB-H IPDC, and DVB-IPTV \cite{MBMS_standard, DVB_standard}.} In FR schemes, the transmitter sends its messages using fixed-length codewords, with retransmissions in case of decoding failures, i.e., without adding IR bits. HARQ schemes feature an initial transmission followed by subsequent transmissions (of possibly varying lengths) of IR that are guided by feedback from the receiver to the transmitter, but not necessarily at a granularity of a single bit as in IIR. One main theme in the findings of works \cite{najm-age-mg11-harq, sac-age-mg1-harq, simeone-age-finite-code, yates-age-erase-code, baknina-age-coding, feng-age-rateless-codes, najm-age-erasure-coding, javani-aoi-erasure, parag-age-coding, ceran-age-harq, arafa-aoi-coding, huang-estimation-harq-control, chen-aoi-coding-bc, feng-coding-aoi-bc} is that optimal codes should strike a balance between using long codewords to minimize channel errors and using short ones to minimize age. Our work in this paper primarily focuses on evaluating the performances of using IIR and FR coding schemes. However, different from all the works in \cite{najm-age-mg11-harq, sac-age-mg1-harq, simeone-age-finite-code, yates-age-erase-code, baknina-age-coding, feng-age-rateless-codes, najm-age-erasure-coding, javani-aoi-erasure}, {\it we consider the additional presence of fixed non-zero receiver processing times.}

The second line of research related to this work is related to evaluating the role of AoI in remote estimation, e.g., \cite{klugel2019aoi-fr, mitra-estimation-graphs-aoi, chakravorty-estimation-pckt-drop-markov, ayan-aoi-voi-cntrl, roth-mse-aoi-finite-blocklength, ramirez-aoi-compression, bastopcu-aoi-distortion, bastopcu-partial-updates, maatouk-aoii, sun-weiner, ornee-aoi-estimation-ou}. The works in \cite{klugel2019aoi-fr, mitra-estimation-graphs-aoi, chakravorty-estimation-pckt-drop-markov} characterize implicit and explicit relationships between mean square error (MSE) and AoI under different estimation contexts; references \cite{ayan-aoi-voi-cntrl, roth-mse-aoi-finite-blocklength} consider the notion of the value of information (mainly through MSE) and show that optimizing it can be different from optimizing AoI; lossy source coding and distorted updates for AoI minimization is considered in \cite{ramirez-aoi-compression, bastopcu-aoi-distortion, bastopcu-partial-updates}; reference \cite{maatouk-aoii} adds more context to AoI by introducing and analyzing a variant metric termed the age of incorrect information (AoII) to capture error in updates; while the works in \cite{sun-weiner, ornee-aoi-estimation-ou} consider sampling of Wiener and OU processes for the purpose of remote estimation, and draw connections between MSE and AoI. Our work in this paper also focuses on characterizing the relationship of MSE and AoI, {\it yet with the additional presence of quantization errors.} It is worth noting that while studying optimal sampling with distortion guarantees is a classical problem, it has been recently approached differently in \cite{kipnis2018a2d}, which characterizes the minimal sampling frequency required to achieve Shannon's rate-distortion function, and concludes that sub-Nyquist sampling can attain the fundamental rate-distortion tradeoff if the energy spectral density of the signal is non-uniform (see \cite{kipnis2018a2d} and the references therein).

\begin{figure}[t]
\center
\includegraphics[scale=.375]{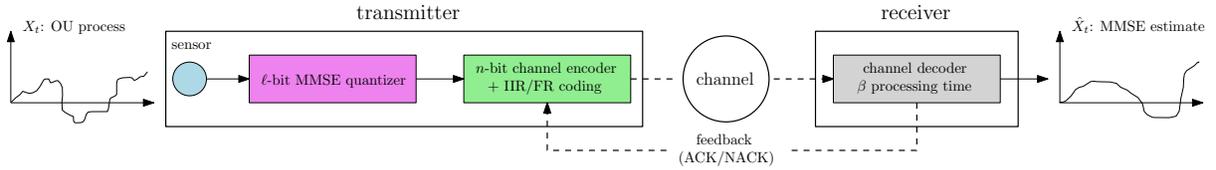}
\caption{System model considered for sampling, quantizing and encoding an OU process at the transmitter, and reconstructing it at the receiver.}
\label{fig_sys_mod}
\end{figure}

While AoI is a time-based metric that has been originally studied in queuing-theoretic frameworks to assess latency, e.g., \cite{yates_age_1, ephremides_age_random, ephremides_age_management}, it is relatively easier to analyze for process tracking purposes compared to MSE, since AoI only takes the statistics of the communication channel into consideration, unlike MSE that also takes the statistics of the process itself into account to assess the {\it quality} of tracking. Under some assumptions, MSE can be shown to have a very dependent behavior on AoI, and hence, minimizing AoI becomes equivalent to minimizing MSE. This is one main idea around which this work revolves, and has been the focus of the works in \cite{sun-weiner, ornee-aoi-estimation-ou}, which are the most closely-related works to ours. References \cite{sun-weiner, ornee-aoi-estimation-ou} derive optimal sampling methods to minimize the long-term average MMSE for Wiener \cite{sun-weiner} and OU \cite{ornee-aoi-estimation-ou} processes. In both works, the communication channel introduces random delays, before perfect (distortion-free) samples are received. It is shown that if sampling times are independent of the instantaneous values of the process (signal-independent sampling) the MMSE reduces to AoI in case of Wiener \cite{sun-weiner}, and to an increasing functional of AoI (age-penalty) in case of OU \cite{ornee-aoi-estimation-ou}. It is then shown that the optimal sampling policy has a threshold structure, in which a new sample is acquired only if the expected AoI in case of Wiener (or age-penalty in case of OU) surpasses a certain value. In addition, signal-dependent optimal sampling policies are also derived \cite{sun-weiner, ornee-aoi-estimation-ou}.

In this work, we consider the transmission of quantized and coded samples of an OU process through a noisy channel. We note that we consider an OU process in our study since, unlike the conventional Wiener process, it has a bounded variance, leading to bounded quantization error as well. Different from \cite{ornee-aoi-estimation-ou}, not every sample has guaranteed reception, and received samples suffer from quantization noise. The receiver uses the received samples to construct an MMSE estimate for the OU process. Quantization and coding introduce a tradeoff: {\it few quantization levels and codeword bits would transmit samples faster, yet with high distortion and probability of error.} An optimal choice, therefore, needs to be made, which depends mainly on how fast the OU process varies as well as the channel errors. Different from related works, effects of having (fixed) {\it non-zero receiver processing times}, mainly due to decoding and sending feedback, are also considered in this work.

We focus on signal-independent sampling, together with an MMSE quantizer, combined with either IIR or FR coding schemes; see Fig.~\ref{fig_sys_mod}. The MMSE of the OU process is first shown to be an increasing functional of AoI, as in \cite{ornee-aoi-estimation-ou}, parameterized directly by the number of quantization bits $\ell$, and indirectly by the number of codeword bits $n$ and the receiver processing time $\beta$. We formulate two problems, one for IIR and another for FR, to choose sampling times so that the long-term average MMSE is minimized. Focusing on stationary deterministic policies, we present optimal solutions for both problems in the case of {\it general increasing age-penalties,} not necessarily corresponding to MMSE, which may be useful in other contexts in which IIR and FR coding schemes are employed. The solution for IIR has a {\it threshold} structure, as in \cite{sun-cyr-aoi-non-linear, ornee-aoi-estimation-ou}, while that for FR is a {\it just-in-time} sampling policy that does not require receiver feedback.

We then present what we call {\it enhanced} IIR and FR schemes, in which we leverage the processing time to our favor through fine-tuning sampling and/or transmission times in such a way that {\it the receiver never waits for data when necessary.} This allows us to mitigate the negative effects of processing times to the most extent possible, and produce timely estimates that are able to track the OU process better. We finally discuss how to select $\ell$ and $n$, and show that the relatively simpler FR scheme can outperform IIR for relatively large values of $\beta$.

The proposed joint optimization of sampling, quantization and coding in this paper takes a step towards achieving the notion of timely real-time tracking of random processes, which can be applied in applications of communications, networks and control. We summarize our main contributions as follows:
\begin{itemize}
\item Presenting a thorough analysis of the effects of quantization and coding (with specific focus on IIR and FR) on the estimation error of Gauss-Markov processes (with specific focus on the OU process). Specifically, we show that there is an inherent relationship between the number of quantization levels and codeword lengths used to convey the samples and the OU process statistics (in particular how fast it varies over time).
\item Characterizing the optimal (signal-independent) sampling strategy (MSE-minimal) for IIR and FR in this context.
\item Introducing, {\it for the first time in the AoI literature (to the best of our knowledge),} the effects of non-zero processing delays at the receiver (for decoding and sending feedback), based on which we argue that one can enhance the performance of both IIR and FR by carefully tailoring the transmission times to the processing delays.
\item Validating our theoretical results by conducting multiple numerical studies and presenting examples that show the effects of the OU process statistics on the optimal quantization levels and coding lengths.
\end{itemize}

Compared to the conference version \cite{arafa-aoi-estimate-ou}, this paper adds: (1) novel thorough analyses of the enhanced schemes in Section~\ref{sec-enhanced}; (2) complete proofs for results and formulas that were omitted in \cite{arafa-aoi-estimate-ou}; and (3) multiple numerical studies that showcase the main results and intuitions.

\section{System Model}

\subsection{Quantization and Coding of the OU Process}

We consider a sensor that acquires time-stamped samples from an OU process. Given a value of $X_s$ at time $s$, the OU process evolves as follows \cite{ou-brownian-motion, doob-brownian-motion}:
\begin{align} \label{eq_ou_evol}
X_t=X_se^{-\theta(t-s)}+\frac{\sigma}{\sqrt{2\theta}}e^{-\theta(t-s)}W_{e^{2\theta(t-s)}-1},\quad t\geq s,
\end{align}
where $W_t$ denotes a Wiener process, while $\theta>0$ and $\sigma>0$ are fixed parameters. The sensor acquires the $i$th sample at time $S_i$ and feeds it to an MMSE quantizer that produces an $\ell$-bit message ready for encoding. We will use the term {\it message} to refer to a quantized sample of the OU process. Let $\tilde{X}_{S_i}$ represent the quantized version of the sample $X_{S_i}$, and let $Q_{S_i}$ denote the corresponding quantization error. Thus,
\begin{align} \label{eq_qntz_smpl}
X_{S_i}=\tilde{X}_{S_i}+Q_{S_i}.
\end{align}
Each message is encoded and sent over a noisy channel to the receiver. The receiver updates an MMSE estimate of the OU process if decoding is successful. ACKs and NACKs are fed back following each decoding attempt. A fixed receiver processing time $\beta$ time units is incurred per each decoding attempt, which also includes the time to generate and send feedback. Channel errors are independent and identically distributed (i.i.d.) across time/messages.

Two channel coding schemes are investigated. The first is IIR, in which a message transmission starts with an $n$-bit codeword, $n\geq\ell$, and then incremental redundancy (IR) bits are added one-by-one if a NACK is received until the message is eventually decoded and an ACK is fed back. The second scheme is FR, in which a message is encoded into fixed $n$-bit codewords, yet following a NACK the message in transmission is discarded and a {\it new} sample is acquired and used instead. Following ACKs, the transmitter may idly wait before acquiring a new sample and sending a new message.\footnote{The main reason behind waiting, as will be shown in details in the sequel, is that it leads to sending fresher samples, which can be more rewarding in terms of the {\it long-term average} MMSE, and {\it not} the instantaneous MMSE. Note that waiting policies have been generously used in previous works that focus on minimizing average AoI, e.g., \cite{yates_age_eh, sun-age-mdp, arafa-age-online-finite}.}

\subsection{Communication Channel}

Let $D_i$ denote the reception time of the $i$th {\it successfully decoded} message. For the IIR scheme, each message is eventually decoded, and therefore
\begin{align}
D_i=S_i+Y_i
\end{align}
for some random variable $Y_i$ that represents the channel delay incurred due to the IR bits added. Let $T_b$ denote the time units consumed per bit transmission. Hence,
\begin{align}
Y_i=nT_b+\beta+r_i(T_b+\beta),
\end{align}
where the random variable $r_i\in\{0,1,2,\dots\}$ denotes the number of IR bits used until the $i$th message is decoded. Note that in the IIR scheme $\beta$ is added for the original $n$-bit codeword transmission, and then for each IR transmission until successful decoding. Let 
\begin{align}
\bar{n}\triangleq nT_b+\beta
\end{align}
for conciseness. Channel delays $Y_i$'s are i.i.d. $\sim Y$, where
\begin{align}
\mathbb{P}\left(Y=\bar{n}\right)=&p_0, \label{eq_Y_dist_1} \\
\mathbb{P}\left(Y=\bar{n}+k(T_b+\beta)\right)=&\prod_{j=0}^{k-1}(1-p_j)p_k,\quad k\geq1, \label{eq_Y_dist_2}
\end{align}
with $p_j$ denoting the probability that an ACK is received when $r_i=j$. This depends on the channel code being used, and the model of the channel errors, yet it holds that $p_j\leq p_{j+1}$.

For the FR scheme, there can possibly be a number of transmission {\it attempts} before a message is eventually decoded. Let $M_i$ denote the number of these attempts in between the $(i-1)$th and $i$th successfully decoded messages, and let $S_{i,j}$ denote the sampling time pertaining to the $j$th attempt of which, $1\leq j\leq M_i$. Therefore, only the $M_i$th message is successfully decoded, and the rest are all discarded. Since each message is encoded using fixed $n$-bit codewords, we have
\begin{align}
D_i=S_{i,M_i}+\bar{n}, \quad\forall i.
\end{align}
Observe that in the FR scheme each successfully-decoded message incurs only {\it one} $\beta$, since each decoding attempt occurs on a message pertaining to a {\it fresh} sample. According to the notation developed for the IIR channel delays above, $M_i$'s are i.i.d. geometric random variables with parameter $p_0$.

\subsection{MMSE Estimation and AoI}

Based on the above notation so far, the AoI at time $t$ is mathematically defined as follows:
\begin{align} \label{eq:aoi-def}
\text{AoI}(t)\triangleq t-u_i(t),\quad D_i\leq t<D_{i+1},
\end{align}
where $u_i(t)$ denotes the time stamp of the latest received sample before time $t$. Thus, for $D_i\leq t<D_{i+1}$, we have $u_i(t)=S_i$ for the IIR scheme, and $u_i(t)=S_{i,M_i}$ for the FR scheme.

Upon successfully decoding a message at time $D_i$, the receiver constructs an MMSE estimate for the OU process. For the purpose of real-time tracking, do not allow retroactive reconstruction of the process, and restrict our attention to MMSE estimators that only use the latest-received information.\footnote{Note that the OU process is no longer Markov after quantization. The implication of this is that the MMSE estimator in (\ref{eq:mmse_estimate_iir}) is potentially suboptimal since it focuses only on the latest received sample. It is, however, simple enough in practice, and admits the analytical solutions derived in the paper. Deriving an optimal MMSE estimator, or showing that considering only the latest received quantized sample performs well enough, e.g., close to optimal, is to be pursued in future work.} For the IIR scheme this is
\begin{align} \label{eq:mmse_estimate_iir}
\hat{X}_t=\mathbb{E}\left[X_t\Big|S_i,\tilde{X}_{S_i}\right],\quad D_i\leq t<D_{i+1}.
\end{align}
Using (\ref{eq_ou_evol}) and (\ref{eq_qntz_smpl}), we have
\begin{align}
\hat{X}_t=&\mathbb{E}\bigg[\tilde{X}_{S_i}e^{-\theta
\left(t-S_i\right)}+Q_{S_i}e^{-\theta\left(t-S_i\right)}+\frac{\sigma}{\sqrt{2\theta}}e^{-\theta\left(t-S_i\right)}W_{e^{2\theta\left(t-S_i\right)}-1}\bigg|S_i,\tilde{X}_{S_i}\bigg] \\
=&\tilde{X}_{S_i}e^{-\theta\left(t-S_i\right)},\quad D_i\leq t<D_{i+1},
\end{align}
where the last equality follows by independence of the Wiener noise in $[D_i,t]$ from $(S_i,\tilde{X}_{S_i})$, and that for the MMSE quantizer, the quantization error is zero-mean \cite{cover}. The MMSE is now given as follows for $D_i\leq t<D_{i+1}$: 
\begin{align}
\texttt{mse}\left(t,S_i\right)=&\mathbb{E}\left[\left(X_t-\hat{X}_t\right)^2\right] \\
=&\mathbb{E}\left[Q_{S_i}^2\right]e^{-2\theta\left(t-S_i\right)}+\frac{\sigma^2}{2\theta}\left(1-e^{-2\theta\left(t-S_i\right)}\right). \label{eq_mse_qnt_dly}
\end{align}
{\it We see from the above that even if $D_i-S_i=0$, i.e., if the $i$th sample is transmitted and received instantaneously, the MMSE estimate at $t=D_i$ would still suffer from quantization errors.} 

In the sequel, we consider $X_0=0$ without loss of generality, and hence, using (\ref{eq_ou_evol}), the variance of $X_t$ is given by $\mathbb{E}\left[X_t^2\right]=\frac{\sigma^2}{2\theta}\left(1-e^{-2\theta t}\right),~t>0$.
Following a rate-distortion approach (note that $X_t$ is Gaussian), the following relates the number of bits $\ell$ and the instantaneous mean square quantization error \cite{cover}:\footnote{There are other works in the literature that consider different kinds of (practical) quantizers and study their effects on filtering stationary Gaussian processes, see, e.g., the uniform quantizer treatment in \cite{poor-84-unfrm-quntzr-Gauss}, which may lead to different quantization errors statistics. Our setting naturally focuses on MMSE quantizers, which are relevant to the purpose of MMSE estimation.}
\begin{align} \label{eq_quant_error_t}
\mathbb{E}\left[Q_t^2\right]=\frac{\sigma^2}{2\theta}\left(1-e^{-2\theta t}\right)2^{-2\ell}, \quad t>0.
\end{align}
Using the above in (\ref{eq_mse_qnt_dly}) and rearranging, we get that
\begin{align}
\!\!\!\texttt{mse}\!\left(t,S_i\right)\!=&\frac{\sigma^2}{2\theta}\!\left(\!1\!-\!\left(1\!-\!2^{-2\ell}\!\left(1\!-\!e^{-2\theta S_i}\right)\right)\!e^{-2\theta\left(t-S_i\right)}\!\right),
\end{align}
We note that as $\ell\rightarrow\infty$, the above expression becomes the same as that derived for the signal-independent sampling scheme analyzed in \cite{ornee-aoi-estimation-ou}. However, since we consider practical coding aspects in this work, as $\ell\rightarrow\infty$, it holds that $n\rightarrow\infty$ as well and no sample will be received.

We focus on dealing with the system in {\it steady state,} in which both $t$ and $S_i$ are relatively large. In this case, the mean square quantization error in (\ref{eq_quant_error_t}) becomes independent of time, and only dependent upon the steady state variance of the OU process $\sigma^2/2\theta$.\footnote{Equivalently, one can initiate the OU process by $X_0\sim\mathcal{N}\left(0,\frac{\sigma^2}{2\theta}\right)$, whence $\mathbb{E}\left[X_t^2\right]=\frac{\sigma^2}{2\theta},~\forall t$.} Hence, in steady state, the MMSE becomes
\begin{align}
\texttt{mse}\left(t,S_i\right)=&\frac{\sigma^2}{2\theta}\left(1-\left(1-2^{-2\ell}\right)e^{-2\theta\left(t-S_i\right)}\right) \\
\triangleq&h_\ell\left(t-S_i\right), \quad D_i\leq t<D_{i+1}, \label{eq_mmse_iir}
\end{align}
which is an increasing functional of the AoI $t-S_i$ in (\ref{eq:aoi-def}). One can see from the MMSE expression above that there exists a tension between the number quantization levels and AoI. In particular, as $\ell$ increases, the quantization noise decreases and the samples transmitted become more precise. However, this necessiates using a larger codeword length $n$, which in turn increases the age-penalty. Hence, a tradeoff exists between sending slow but precise samples and fast but less accurate ones. We discuss how to optimally characterize this tradeoff in Section~\ref{sec_cmpr_iir_fr}.

For the FR scheme, the analysis follows similarly, after adding one more random variable denoting the number of transmissions, $M_i$. Specifically, it holds that
\begin{align}
\hat{X}_t=&\tilde{X}_{S_{i,M_i}}e^{-\theta\left(t-S_{i,M_i}\right)}, \\
\texttt{mse}\left(t,S_{i,M_i}\right)=&h_\ell\left(t-S_{i,M_i}\right), \quad D_i\leq t<D_{i+1}. \label{eq_mmse_fr}
\end{align}

We see from (\ref{eq_mmse_iir}) and (\ref{eq_mmse_fr}) that there are two main contributing factors to the MMSE. The first is due to quantization, represented by the factor $\left(1-2^{-2\ell}\right)$, and the second is due to the channel delay, added mainly because of coding and errors, represented by the AoI $t-S$.

\section{Optimal Sampling Policies: General Age-Penalty}

The main goal is to choose the sampling times, for given $\ell$, $n$ and $\beta$, such that the long-term average MMSE is minimized. In this section, we formulate two problems to achieve such goal: one for IIR and another for FR, and present their solutions in the upcoming section. Later on in Section~\ref{sec_cmpr_iir_fr}, we discuss how to choose the best $\ell$ and $n$, as well as compare the performances of IIR and FR in general.

For both coding schemes, let us denote by an {\it epoch} the time elapsed in between two successfully received messages. Thus, the $i$th epoch starts at $D_{i-1}$ and ends at $D_i$, with $D_0\equiv0$.

\begin{remark}
Our analysis does not depend on the specific structure of the MMSE functional $h_\ell(\cdot)$; it extends to any differentiable increasing age-penalty functional $g(\cdot)$. Therefore, in what follows, we formulate our problems and present their solutions for the case of minimizing a long-term average age-penalty, making the results applicable in other contexts.
\end{remark}

\subsection{The IIR Scheme}

For the IIR scheme, the problem is formulated as
\begin{align} \label{opt_main_iir}
\min_{\{S_i\}}\quad\limsup_{l\rightarrow\infty}\frac{\sum_{i=0}^l\mathbb{E}\left[\int_{D_i}^{D_{i+1}}g\left(t-S_i\right)dt\right]}{\sum_{i=0}^l\mathbb{E}\left[D_{i+1}-D_i\right]},
\end{align}
where the numerator represents the total age-penalty (the MMSE in case of the OU process estimation) across all epochs, and the denominator represents the total time.

Let us define $W_i$ as the waiting time at the beginning of the $i$th epoch before acquiring the $i$th sample. That is, $S_i=D_{i-1}+W_i$. Therefore, one can equivalently solve for the waiting times $W_i$'s instead of sampling times $S_i$'s. We focus on a class of {\it stationary deterministic} policies in which
\begin{align}
W_i=f\left(g\left(D_{i-1}-S_{i-1}\right)\right),\quad\forall i.
\end{align}
That is, {\it the waiting time in the $i$th epoch is a deterministic function of its starting age-penalty value.} Such focus is motivated by the fact that channel errors are i.i.d. and by its optimality in similar frameworks, e.g., \cite{sun-age-mdp, jing-age-online, arafa-age-online-finite}. Defining $w\triangleq f\circ g$ and noting that $D_{i-1}-S_{i-1}=Y_{i-1}$ we have
\begin{align}
W_i=w\left(Y_{i-1}\right),
\end{align}
which induces a stationary distribution of $D_i$'s and the age-penalty across all epochs. Due to stationarity, we can now drop the epoch's index $i$, and (re)define notations used in a typical epoch. It starts at time $\overline{D}$ with AoI $\overline{Y}$, and with the latest sample acquired at time $\overline{S}$, such that $\overline{D}=\overline{S}+\overline{Y}$. Then, a waiting time of $w\left(\overline{Y}\right)$ follows, after which a new sample is acquired, quantized, and transmitted, taking $Y$ time units to reach the receiver at time $D=\overline{D}+w\left(\overline{Y}\right)+Y$, which is the epoch's end time. Therefore, problem (\ref{opt_main_iir}) now reduces to a minimization over a single epoch as follows:
\begin{align} \label{opt_iir_epoch}
\min_{w(\cdot)\geq0}\quad\frac{\mathbb{E}\left[\int_{\overline{D}}^{\overline{D}+w\left(\overline{Y}\right)+Y}g\left(t-\overline{S}\right)dt\right]}{\mathbb{E}\left[w\left(\overline{Y}\right)+Y\right]}.
\end{align}
Given the realization of $\overline{Y}$ at time $\overline{D}$, the transmitter decides on the waiting time $w\left(\overline{Y}\right)$ that minimizes the long-term average age-penalty demonstrated in the fractional program above.\footnote{We now see explicitly how waiting can be beneficial. Since waiting increases {\it both} the numerator and denominator of the objective function of problem (\ref{opt_iir_epoch}), its optimal value can be non-zero.} 

We follow Dinkelbach's approach to transform (\ref{opt_iir_epoch}) into the following auxiliary problem for fixed $\lambda\geq0$ \cite{dinkelbach-fractional-prog}:
\begin{align} \label{opt_iir_aux}
p^{IIR}(\lambda)\triangleq\min_{w(\cdot)\geq0}\quad\mathbb{E}\left[\int_{\overline{D}}^{\overline{D}+w\left(\overline{Y}\right)+Y}g\left(t-\overline{S}\right)dt\right] -\lambda\mathbb{E}\left[w\left(\overline{Y}\right)+Y\right].
\end{align}
The optimal solution of (\ref{opt_iir_epoch}) is then given by $\lambda^*_{IIR}$ that solves $p^{IIR}(\lambda^*_{IIR})=0$, which can be found via bisection, since $p^{IIR}(\lambda)$ is decreasing \cite{dinkelbach-fractional-prog}. The following theorem characterizes the solution of problem (\ref{opt_iir_aux}). The proof is in Appendix~\ref{apndx_pf_iir_main_result}.

\begin{theorem} \label{thm_iir_main_result}
The optimal solution of problem (\ref{opt_iir_aux}) is given by
\begin{align}
w^*(\bar{y})=\left[G_{\bar{y}}^{-1}(\lambda)\right]^+,
\end{align}
where $\left[\cdot\right]^+\triangleq\max(\cdot,0)$, $\bar{y}$ is the realization of the starting AoI $\bar{Y}$, and $G_{\bar{y}}(x)\triangleq\mathbb{E}\left[g\left(\bar{y}+x+Y\right)\right]$.
\end{theorem}

We note that the theorem can be shown using the result reported in \cite[Theorem~1]{sun-cyr-aoi-non-linear}. Our proof approach, however, is different, and is reported here for completeness. Such approach is also used to show parts of Theorem~\ref{thm_fr_main_result} below.

The optimal waiting policy for IIR has a {\it threshold} structure: a new sample is acquired only when the expected age-penalty by the end of the epoch is at least $\lambda$. Note that the optimal $\lambda^*_{IIR}$ corresponds to the optimal long-term average age-penalty.

\subsection{The FR Scheme}

For the FR scheme, the formulated problem can be derived similarly, {\it with the addition of possible waiting times in between retransmissions.}\footnote{This is only amenable for FR since waiting leads to acquiring a fresher sample, and possibly reduced age-penalties. For IIR, waiting after a NACK is clearly suboptimal since it elongates the channel delay for the {\it same} sample.}
Specifically, let $W_{i,j}$ represent the waiting time before the $j$th transmission attempt in the $i$th epoch. A stationary deterministic policy\footnote{We note that \cite{klugel2019aoi-fr} shows the optimality of stationary policies in a time-slotted system in which samples are conveyed through an erasure channel. This resembles our FR model yet with no quantization or coding.} here is such that $W_{i,j}$ is a determinisitc function $w(\cdot)$ of the instantaneous age-penalty. This makes the waiting time before the first transmission attempt given by
\begin{align}
W_{i,1}=&f\left(g\left(D_{i-1}-S_{i-1,M_{i-1}}\right)\right)=w\left(\bar{n}\right)\equiv w_1,
\end{align}
where $D_{i-1}-S_{i-1,M_{i-1}}=\bar{n}$ represents the starting AoI of the $i$th (and every) epoch, following $M_{i-1}$ transmission attempts in the previous one. The waiting time before the second attempt, if needed, will then be given by
\begin{align}
W_{i,2}=&w\left(\bar{n}+w_1+\bar{n}\right)\equiv w_2,
\end{align}
since the AoI before the second attempt is given by the starting AoI of the epoch in addition to the time needed to finish the first transmission attempt. In general, the waiting time before the $j$th attempt in the epoch is given by
\begin{align}
W_{i,j}=&w\left(\sum_{l=1}^{j-1}w_l+j\bar{n}\right)\equiv w_j,
\end{align}
and so on. Therefore, under the FR scheme, a stationary deterministic policy reduces to a countable sequence $\{w_j\}$. 

Proceeding with the same notations for a given epoch as in the IIR scheme, let us define $M$ as the number of transmission attempts in the epoch, $\bar{M}$ as those in the previous epoch, and $\overline{S}_{\bar{M}}$ as the sampling time of the successful (last) transmission attempt in the previous epoch. The problem now becomes
\begin{align} \label{opt_fr_epoch}
\min_{\{w_j\geq0\}} \quad \frac{\mathbb{E}\left[\int_{\overline{D}}^{\overline{D}+\sum_{j=1}^Mw_j+M\bar{n}}g\left(t-\overline{S}_{\bar{M}}\right)dt\right]}{\mathbb{E}\left[\sum_{j=1}^Mw_j+M\bar{n}\right]}.
\end{align}

We follow a similar approach here as in the IIR scheme and consider the following auxiliary problem:
\begin{align} \label{opt_fr_aux}
p^{FR}(\lambda)\!\triangleq\!\min_{\{w_j\geq0\}} \mathbb{E}\left[\int_{\overline{D}}^{\overline{D}+\sum_{j=1}^Mw_j+M\bar{n}}g\left(t-\overline{S}_{\bar{M}}\right)dt\right]-\lambda\mathbb{E}\left[\sum_{j=1}^Mw_j+M\bar{n}\right].
\end{align}
The optimal solution of problem (\ref{opt_fr_epoch}) is now given by $\lambda^*_{FR}$ that solves $p^{FR}\left(\lambda^*_{FR}\right)=0$, which we will actually provide in {\it closed-form} this time. The optimal waiting policy structure is provided in the next theorem. The proof is in Appendix~\ref{apndx_pf_fr_main_result}. 

\begin{theorem} \label{thm_fr_main_result}
The optimal solution of problem (\ref{opt_fr_aux}) is given by
\begin{align}
w_1^*=&\left[G^{-1}(\lambda)\right]^+, \\
w_j^*=&0,~j\geq2,
\end{align}
where $G(x)\triangleq\mathbb{E}\left[g\left(\bar{n}+x+M\bar{n}\right)\right]$. In addition, the optimal solution of problem (\ref{opt_fr_epoch}), $\lambda^*_{FR}$, is such that $w_1^*=\left[G^{-1}\left(\lambda^*_{FR}\right)\right]^+=0$.
\end{theorem}

A closed-form expression for $\lambda^*_{FR}$ can now be found via substituting $w_j=0,~\forall j$ in (\ref{opt_fr_epoch}). 

Theorem~\ref{thm_fr_main_result} shows that {\it zero-wait} policies are optimal for FR, which is quite intuitive. First, waiting is not optimal in between retransmissions, even though it would lead to acquiring fresher samples, since the AoI is already relatively high following failures. Second, since the epoch always starts with the same AoI, $\bar{n}$, one can optimize the long-term average age-penalty to make waiting not optimal at the beginning of the epoch as well. We note, however, that such results do {\it not} follow from \cite[Theorem~5]{sun-age-mdp}, since there can be multiple transmissions in the same epoch. We also note that while zero-wait policies have been invoked in other works involving FR coding schemes, e.g., \cite{yates-age-erase-code, najm-age-erasure-coding}, Theorem~\ref{thm_fr_main_result} provides a proof of their {\it optimality} for general increasing age-penalties. Finally, we note that the results of Theorem~\ref{thm_fr_main_result} are related to those reported in Propositions 3 and 6 in \cite{klugel2019aoi-fr}. However, our proof of the optimality of the threshold policy is based on a quite different Lagrangian approach that works for continuous-time systems (different from the time-slotted system considered in \cite{klugel2019aoi-fr}).

\section{Enhanced Transmission Schemes} \label{sec-enhanced}

So far the analysis assumed that, naturally, the transmitter must wait for feedback before taking new decisions, e.g., sending IR bits in case of the IIR scheme or acquiring a new sample in case of the FR scheme. In this section, we show that such waiting for receiver processing is unnecessary. We basically take advantage of the processing time $\beta$ to send extra pieces of information when possible, in order to maintain a smooth information supply {\it as the receiver decodes and processes previous messages.} We show that with proper timing, this can lead to better results for both the IIR and FR schemes, and hence the name {\it enhanced} schemes. One assumption here is that the receiver has a (possibly-infinite) queue to store unprocessed data.

\subsection{Enhanced IIR Scheme}

The enhanced IIR scheme works as follows. The transmitter sends the original $n$-bit codeword, consuming $nT_b$ time units, after which the receiver starts decoding. Then, instead of waiting for the $\beta$ time units processing time, the transmitter goes ahead with transmitting IR bits continuously. This way, if the original $n$-bit codeword is not successfully decoded, the receiver would have some IR bits awaiting in its queue ready for processing, which saves time. The continuous stream of IR bits transmission stops whenever an ACK is fed back. We note that if the ACK is received in the middle of a bit transmission, this transmission is cut off and stopped immediately.

The next lemma shows that the enhanced IIR scheme described above experiences (almost surely) smaller channel delay for each message transmission. The proof is in Appendix~\ref{apndx_pf_iir_enhanced}.

\begin{lemma} \label{thm_iir_enhanced}
For a given value of $r_i$, the enhanced IIR scheme saves the following amount of time in channel delay during the $i$th epoch:
\begin{align} \label{eq_iir_enhanced_delay}
r_i\min\{\beta,T_b\}+(r_i-\kappa_i)\beta \cdot \mathbbm{1}_{\beta\geq T_b}, 
\end{align}
where $\kappa_i$ is the smallest integer in $\{0,1,\dots,r_i\}$ such that $\lfloor\kappa_i\beta/T_b\rfloor\geq r_i$, with $\lfloor x\rfloor$ denoting the largest integer smaller than or equal to $x$, and $\mathbbm{1}_A=1$ if event $A$ is true and $0$ otherwise.
\end{lemma}

Lemma~\ref{thm_iir_enhanced} shows that the enhanced IIR scheme would achieve smaller long-term average age-penalty relative to the original IIR scheme discussed previously, owing to (\ref{eq_iir_enhanced_delay}). The intuition behind this is that once a new sample is generated, its AoI counter starts to increase, and hence the faster it reaches the destination the better. This is different from idle waiting, however, since the waiting occurs {\it before} the sample is generated.

Let $\tilde{Y}_i$ denote the channel delay experienced by the $i$th message using the enhanced IIR scheme. Such $\tilde{Y}_i$'s are i.i.d.~$\tilde{Y}$. Using the same notation used to describe the distribution of (the original channel delay) $Y$ in (\ref{eq_Y_dist_1}) and (\ref{eq_Y_dist_2}), the enhanced IIR channel delay $\tilde{Y}$ has the following distribution according to Lemma~\ref{thm_iir_enhanced}:
\begin{align}
\mathbb{P}\left(\tilde{Y}=\bar{n}\right)=&p_0, \\
\mathbb{P}\left(\tilde{Y}=\bar{n}+kT_b\right)=&\prod_{j=0}^{k-1}(1-p_j)p_k,\quad k\geq1,
\end{align}
for $\beta<T_b$, and
\begin{align}
\mathbb{P}\left(\tilde{Y}=\bar{n}\right)=&p_0, \\
\mathbb{P}\left(\tilde{Y}=\bar{n}+k\beta\right)=&\prod_{j=0}^{k-1}\left(1-p_{\left\lfloor \frac{(k-1)\beta}{T_b} \right\rfloor}\right)p_{\left\lfloor \frac{k\beta}{T_b} \right\rfloor},\quad k\geq1,
\end{align}
for $\beta\geq T_b$. One would then apply the results of Theorem~\ref{thm_iir_main_result} to find the optimal waiting policy in accordance to the enhanced IIR channel delay distribution $\tilde{Y}$ specified above.

\subsection{Enhanced FR Scheme}

For FR, since zero-waiting is optimal by Theorem~\ref{thm_fr_main_result}, it could be rewarding therefore, age-wise, to send a new message right away after the previous one is {\it delivered}, i.e., after $nT_b$ time units instead of $\bar{n}$. However, this may not be optimal if $\beta$ is relatively large, since it would lead to accumulating {\it stale} messages at the receiver's end as they wait for decoding to finish. 

Let $\delta$ denote the waiting time following a message {\it delivery} at which a new message is transmitted. In the original FR scheme, by Theorem~\ref{thm_fr_main_result}, we had $\delta=\beta$. In general though, $\delta\in[0,\beta]$ and should be optimized. The next lemma provides a solution to the optimal $\delta^*$. The proof is in Appendix~\ref{apndx_pf_fr_enhanced}.

\begin{lemma} \label{thm_fr_enhanced}
In the FR scheme, it is optimal to send a new message after the previous one's delivery by $\delta^*=\left[\beta-nT_b\right]^+$ time units.
\end{lemma}

Lemma~\ref{thm_fr_enhanced} shows that {\it just-in-time} updating is optimal. For $\beta\leq nT_b$, a new sample is acquired and transmitted just-in-time as the previous message is delivered. While for $\beta>nT_b$, a new sample is acquired and transmitted such that it is delivered just-in-time as the receiver finishes decoding the previous message. This way, delivered samples are always fresh, the receiver is never idle, and feedback is unnecessary.

\section{Performance Evaluations and Comparisons} \label{sec_cmpr_iir_fr}

In this section, we discuss how the IIR and FR schemes perform relative to each other under variant system parameters and channel conditions. We do so in the original context of OU process estimation, i.e., when $g(\cdot)\equiv h_\ell(\cdot)$. We note that since the FR scheme has an optimal waiting time of $0$, according to Theorem~\ref{thm_fr_main_result}, it becomes equivalent to a {\it uniform} sampling scheme with fixed sampling frequency that depends on $\ell$, $n$, and $\beta$. In particular, the enhanced FR scheme generates a new sample every $nT_b+\left[\beta-nT_b\right]^+=\max\left\{nT_b,\beta\right\}$ time units. The optimal choice of $\ell$ and $n$, therefore, implicitly provides the optimal (uniform) sampling frequency. Due to the wide use of uniform sampling schemes in practice, the FR scheme serves as an implicit uniform sampling benchmark in our evaluations.

Applying Theorem~\ref{thm_iir_main_result} and Lemma~\ref{thm_iir_enhanced}'s result, the optimal waiting policy for enhanced IIR is
\begin{align} \label{eq_iir_enhanced_wait}
w^*\!\left(\bar{y}\right)\!=\!\left[\frac{1}{2\theta}\log\left(\frac{\frac{\sigma^2}{2\theta}\left(1-2^{-2\ell}\right)\mathbb{E}\left[e^{-2\theta \tilde{Y}}\right]}{\frac{\sigma^2}{2\theta}-\lambda^*_{IIR}}\right)-\bar{y}\right]^+,
\end{align}
where $\tilde{Y}$ is as defined following Lemma~\ref{thm_iir_enhanced}.\footnote{With a slight abuse of notation here, $\bar{y}$ now represents the realization of $\tilde{Y}$ that ended the previous epoch.} In addition, observing that $\frac{\sigma^2}{2\theta}2^{-2\ell}\leq h_{\ell}\left(t-\overline{S}\right)\leq\frac{\sigma^2}{2\theta}$ holds true $\forall t\geq\overline{S}$, one can directly see that $\lambda^*_{IIR}\in\left[2^{-2\ell}\frac{\sigma^2}{2\theta},\frac{\sigma^2}{2\theta}\right]$, facilitating the bisection search. Applying Theorem~\ref{thm_fr_main_result} and Lemma~\ref{thm_fr_enhanced}'s results, the optimal long-term average MMSE for enhanced FR is given by
\begin{align} \label{eq_fr_enhanced_mse}
\frac{\sigma^2}{2\theta}\left(\!1\!-\!\frac{\left(1-2^{-2\ell}\right)e^{-2\theta\bar{n}}p_0}{2\theta K_{n,\beta}}\frac{1-e^{-2\theta K_{n,\beta}}}{1\!-\!(1-p_0)e^{-2\theta K_{n,\beta}}}\right),
\end{align}
where $K_{n,\beta}\triangleq\max\{\beta,nT_b\}$. Derivation details for (\ref{eq_iir_enhanced_wait}) and (\ref{eq_fr_enhanced_mse}) are in Appendix~\ref{apndx_fr_enhanced_mse}.

We consider a binary symmetric channel (BSC) with crossover probability $\epsilon\in\left(0,\frac{1}{2}\right)$, and use maximum distance separable (MDS) codes for transmission. This allows us to write $p_j=\sum_{l=0}^{\lfloor\frac{n+j-\ell}{2}\rfloor}\binom{n+j}{l}\epsilon^l(1-\epsilon)^{n+j-l}$. We set $\sigma^2=1$, and $T_b=0.05$ time units. We refer to enhanced IIR and FR without using the word enhanced throughout this section for convenience.

\subsection{Optimal $(\ell,n)$: Effect of Memory Factor $\theta$}

\begin{figure}[t]
\begin{subfigure}{.5\textwidth}
\center
\includegraphics[scale=.4]{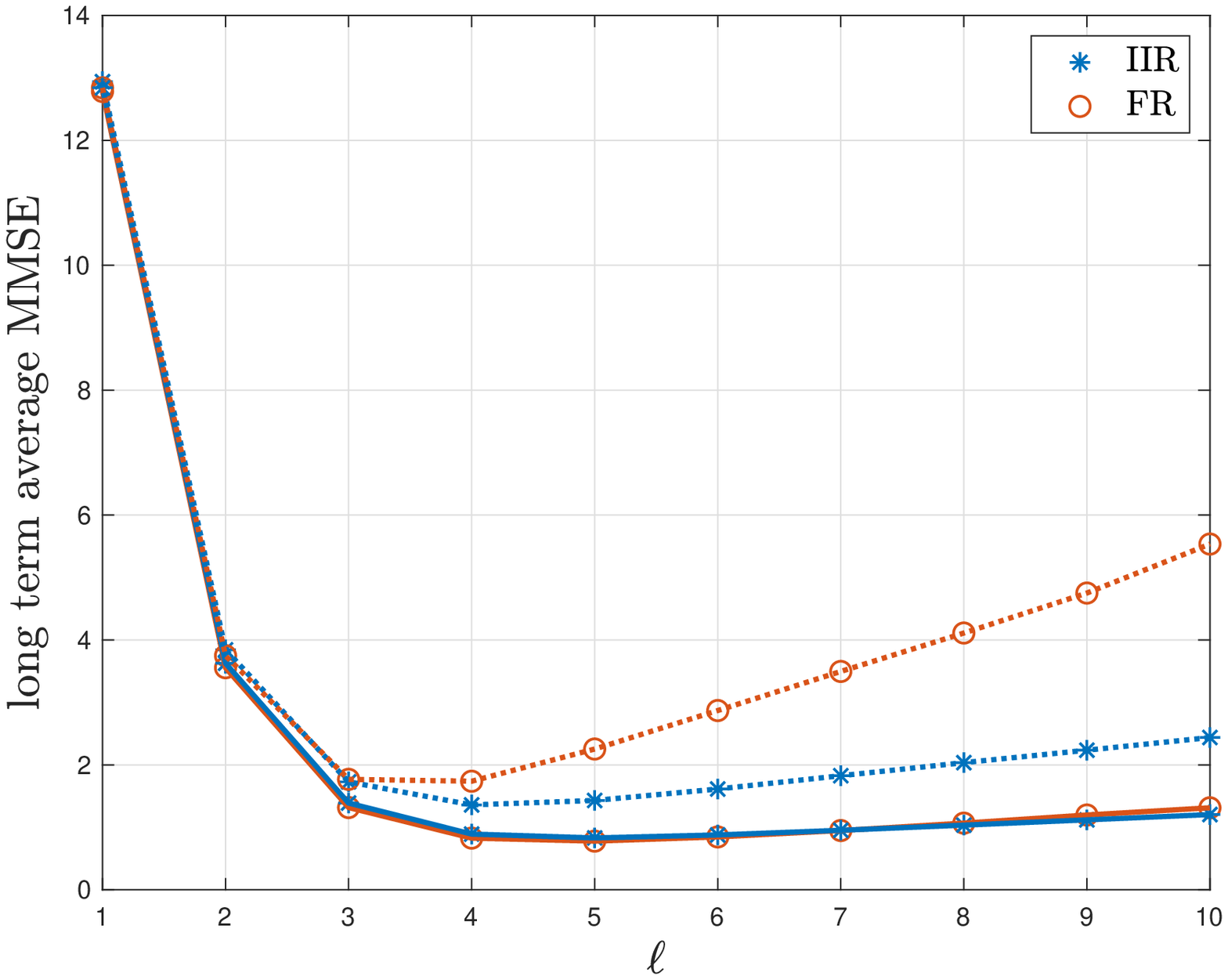}
\caption{$\theta=0.01$}
\label{fig_iir_and_fr_vs_ell_theta_01_beta_15_eps_1_4}
\end{subfigure}
\begin{subfigure}{.5\textwidth}
\center
\includegraphics[scale=.4]{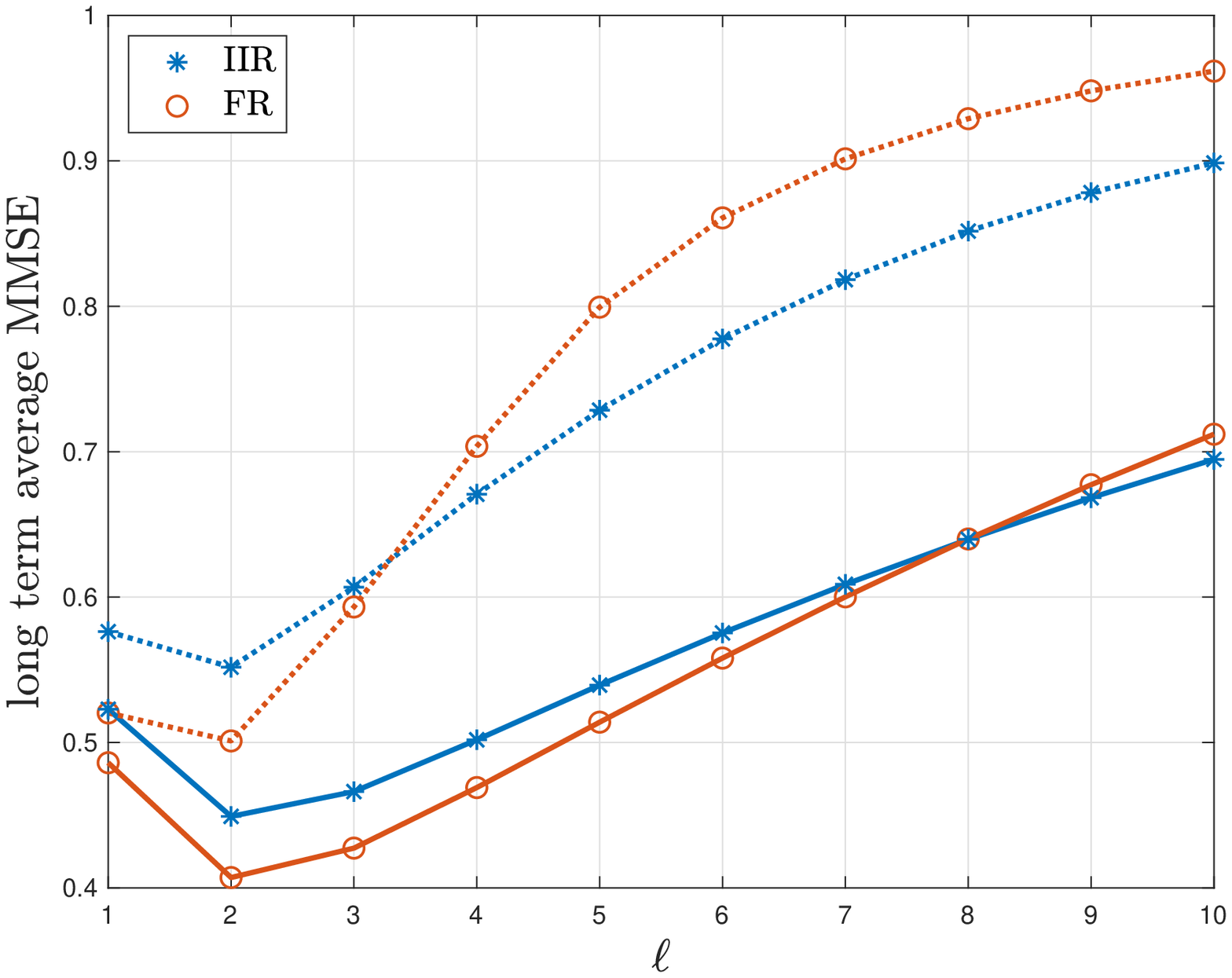}
\caption{$\theta=0.5$}
\label{fig_iir_and_fr_vs_ell_theta_5_beta_15_eps_1_4}
\end{subfigure}
\caption{Performance comparison of IIR and FR vs. $\ell$ for $\beta=0.15$, with $\theta=0.01$ in Fig.~\ref{fig_iir_and_fr_vs_ell_theta_01_beta_15_eps_1_4} (slowly-varying OU process) and $\theta=0.5$ in Fig.~\ref{fig_iir_and_fr_vs_ell_theta_5_beta_15_eps_1_4} (fast-varying OU process). Solid lines: $\epsilon=0.1$, and dotted lines: $\epsilon=0.4$. For $\theta=0.01$, the optimal $(\ell,n)$ pair for both schemes is given by $(5,7)$ for $\epsilon=0.1$ and by $(4,6)$ for $\epsilon=0.4$. While for $\theta=0.5$, the optimal $(\ell,n)$ pair for both schemes is given by $(2,4)$ for both values of $\epsilon$.}
\label{fig_iir_and_fr_vs_ell_beta_15_eps_1_4}
\end{figure}

For fixed $\beta=0.15$, we vary $\ell$ and numerically choose the best $n$ for IIR and FR. We plot the long-term average MMSE for both IIR and FR versus $\ell$ in Fig.~\ref{fig_iir_and_fr_vs_ell_beta_15_eps_1_4}. We do so for $\theta=0.01$ in Fig.~\ref{fig_iir_and_fr_vs_ell_theta_01_beta_15_eps_1_4} (slowly-varying OU process) and $\theta=0.5$ in Fig.~\ref{fig_iir_and_fr_vs_ell_theta_5_beta_15_eps_1_4} (fast-varying OU process). For each value of $\ell$, the optimal $n$ is evaluated. For both values of $\theta$, we repeat the analysis for $\epsilon=0.1$ (in solid lines) and $\epsilon=0.4$ (in dotted lines).

In all of the cases considered, the optimal $n^*=\ell^*+2$. While the optimal $\ell^*$ itself depends on whether the OU processes is slowly ($\theta=0.01$) or fast ($\theta=0.5$) varying. Specifically, we notice that $\ell^*$ decreases with $\theta$. This is intuitive since for slowly-varying processes, one can tolerate larger waiting times to get high quality estimates, and vice versa. It is also shown in the figure that IIR performs better than FR for slowly-varying processes, and vice versa for fast-varying ones. This observation settles a goal that this paper is seeking regarding whether one should send fast low-quality samples or slow high-quality ones for the purpose of remote estimation and tracking; it depends on the memory the time-varying process possesses, abstracted by the variable $\theta$ in this case. We also note that the relationship $n^*=\ell^*+2$ does not always hold, neither it is the case that the optimal $(\ell^*,n^*)$ pairs are the same for IIR and FR; it all depends on the parameters used in the numerical evaluations. If, for instance, we set $\theta=0.01$, $\epsilon=0.4$ and $\beta=1$, we find that the optimal $(\ell^*,n^*)$ pairs are given by $(4,10)$ for IIR, and $(4,18)$ for FR. This can be attributed to the fact that one is estimating a slowly-varying process, over a channel that introduces errors with relatively high rate, with an estimator that incurs a relatively large processing delay ($\beta=20T_b$).

\subsection{IIR vs. FR: Effect of Processing Time $\beta$}

\begin{figure}[t]
\center
\includegraphics[scale=.5]{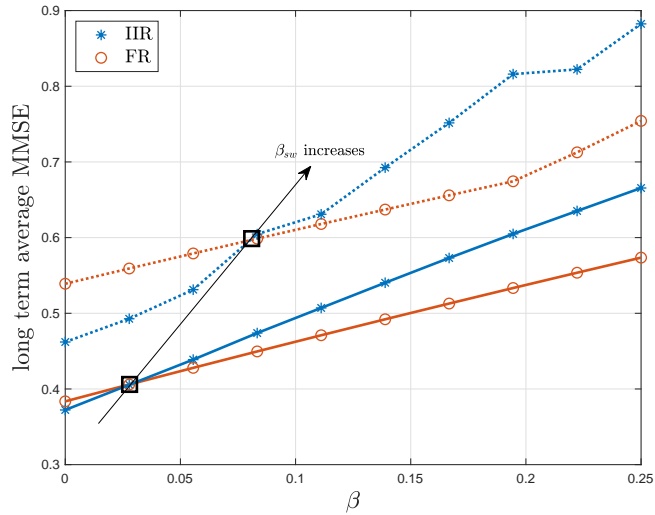}
\caption{Performance comparison of IIR and FR vs. $\beta$, with $\theta=0.25$. Solid lines: $\epsilon=0.1$, and dotted lines: $\epsilon=0.4$. The processing time value after which FR beats IIR, $\beta_{sw}$, is marked in black squares, and is increasing with $\epsilon$.}
\label{fig_iir_and_fr_vs_beta_theta_25_eps_1_4}
\end{figure}

In Fig.~\ref{fig_iir_and_fr_vs_beta_theta_25_eps_1_4}, we fix $\theta=0.25$ and plot the long-term average MMSE achieved by IIR and FR versus $\beta$. We do so for $\epsilon=0.1$ (in solid lines) and $\epsilon=0.4$ (in dotted lines). We observe that IIR performs better than FR for relatively lower values of $\beta$, and then the performance switches after some $\beta_{sw}$ processing time value, marked in black squares. We note that the reason why the curves for $\epsilon=0.4$ are not very smooth is mainly attributed to the $\lfloor\cdot\rfloor$ (floor) function used in the enhanced schemes' channel delay calculations.

We notice that the value of $\beta_{sw}$ increases with $\epsilon$, i.e., when the channel becomes worse. However, the gain due to switching from IIR to FR also increases and becomes more rewarding in this case too. As evident from Figs.~\ref{fig_iir_and_fr_vs_ell_beta_15_eps_1_4} and~\ref{fig_iir_and_fr_vs_beta_theta_25_eps_1_4}, there is no coding scheme that dominantly outperforms the other; it all depends on the system parameters comprising the process, the channel and the processing time.

\subsection{Enhanced vs.~Non-Enhanced Schemes}

\begin{figure}[t]
\begin{subfigure}{.5\textwidth}
\center
\includegraphics[scale=.4]{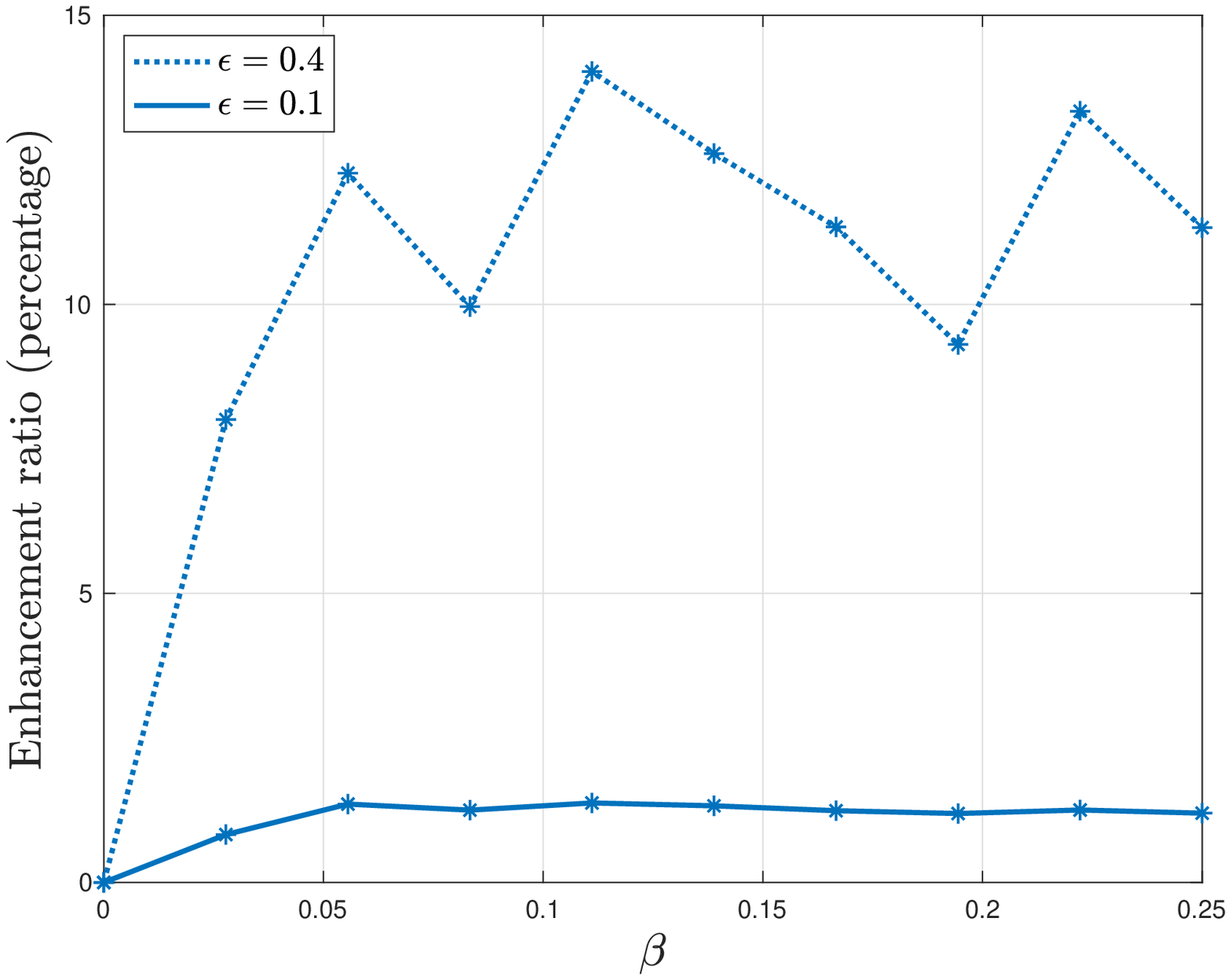}
\caption{IIR}
\label{fig_enhanced_vs_non_IIR_theta_25_eps_1_4}
\end{subfigure}
\begin{subfigure}{.5\textwidth}
\center
\includegraphics[scale=.4]{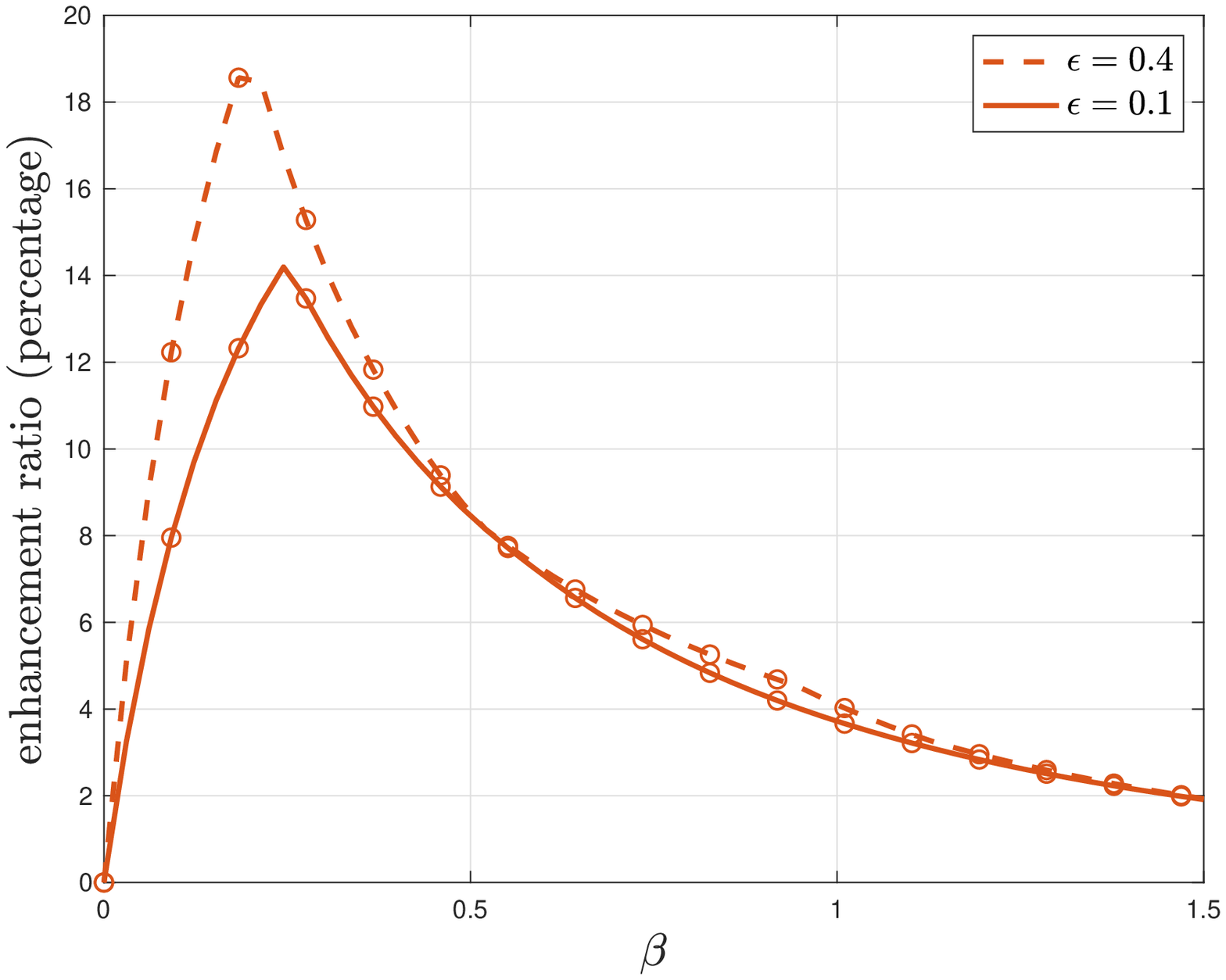}
\caption{FR}
\label{fig_enhanced_vs_non_FR_theta_25_eps_1_4}
\end{subfigure}
\caption{Evaluating the {\it gain} due to enhancement, with $\theta=0.25$. The enhancement ratio is defined as the ratio between the long-term average MMSE of the enhanced scheme to that of the non-enhanced scheme, subtracted from unity.}
\label{fig_enhanced_vs_non_IIR_and_FR_theta_25_eps_1_4}
\end{figure}

We turn our attention to evaluating the gain achieved (i.e., the loss in MMSE) due to employing the enhanced schemes. Specifically, for fixed $\theta=0.25$, let us denote by $\widetilde{\texttt{mmse}}(\beta)$ and $\texttt{mmse}(\beta)$ the long-term average MMSE achieved by the enhanced and the non-enhanced schemes, respectively. We define the enhancement ratio as
\begin{align}
1-\frac{\widetilde{\texttt{mmse}}(\beta)}{\texttt{mmse}(\beta)},
\end{align}
and so the higher this ratio is, the larger the gain due to enhancement. In Fig.~\ref{fig_enhanced_vs_non_IIR_and_FR_theta_25_eps_1_4}, we plot the enhancement ratio (in percentage) for both IIR and FR versus $\beta$.

For the IIR case in Fig.~\ref{fig_enhanced_vs_non_IIR_theta_25_eps_1_4}, we observe that: $(1)$ the enhancement ratio relatively increases with $\beta$ (again, the non-smoothness effect is mainly due to using the floor function in calculations), because as $\beta$ increases, one can {\it fit more data} as the receiver decodes previous ones; and $(2)$ the gain is more apparent for worse channel conditions, which is due to the ability of enhanced IIR to make more data available for reprocessing at the receiver's end following decoding errors, compared to non-enhanced IIR.

Fig.~\ref{fig_enhanced_vs_non_FR_theta_25_eps_1_4} deals with FR, and exhibits some behavioral differences when compared to IIR. In particular, the enhancement ratio first increases then decreases with $\beta$. The reason for such behavior is that for the enhanced FR scheme there is {\it only one} extra codeword that can be transmitted as the receiver finishes processing, {\it regardless} of the value of $\beta$. Specifically, according to Lemma~\ref{thm_fr_enhanced}, the optimal inter-sampling (and transmission) delay for the enhanced FR scheme is given by $\beta-nT_b$ (for $\beta>nT_b$). While for the non-enhanced FR scheme, the inter-sampling delay is given by $\beta$. Hence, as $\beta$ becomes much larger than $nT_b$, the two inter-sampling delays become equivalent, and the performances of both schemes (enhanced and non-enhanced) become similar. Therefore, for the FR scheme, intermediate values of $\beta$ (relative to $nT_b$) provide the highest gain from enhancement. As in the IIR scheme, the enhancement gain is more apparent in worse channel conditions.

In summary, this numerical calculation shows that the enhancement effect is relatively more noticeable for FR ($\approx18\%$ gain) than it is for IIR ($\approx14\%$ gain), and that it would better serve both schemes in relatively worse channel conditions.

\subsection{Timely Real-time Tracking}

We finally apply the techniques developed in this paper to an example sample path of the OU process. In this particular example we fix $\beta=0.15$, $\theta=0.01$ and $\epsilon=0.1$. We first generate an OU process sample path over $t=500$ time units ($10^4\times T_b$). Then, we pass it through an MMSE quantizer\footnote{We train a quantizer using $1000$ different OU processes sample paths, each over $t\in[0,500]$, using Lloyd's algorithm to build this \cite{cover}. Each sample path realization produces a particular code when Lloyd's algorithm converges. We then average over all the produced codes and use the averaged code to generate the results of this subsection.} with $\ell=5$ (which is the optimal $\ell^*$ in this case using Fig.~\ref{fig_iir_and_fr_vs_ell_theta_01_beta_15_eps_1_4}). After that, we use either IIR or FR with $n=7$ (again, this is the optimal $n^*$ in this case) to send the quantized samples through a BSC($0.1$). We apply the optimal waiting policies in accordance to the channel delay realizations and receiver processing time.

\begin{figure}[t]
\begin{subfigure}{.5\textwidth}
\center
\includegraphics[scale=.4]{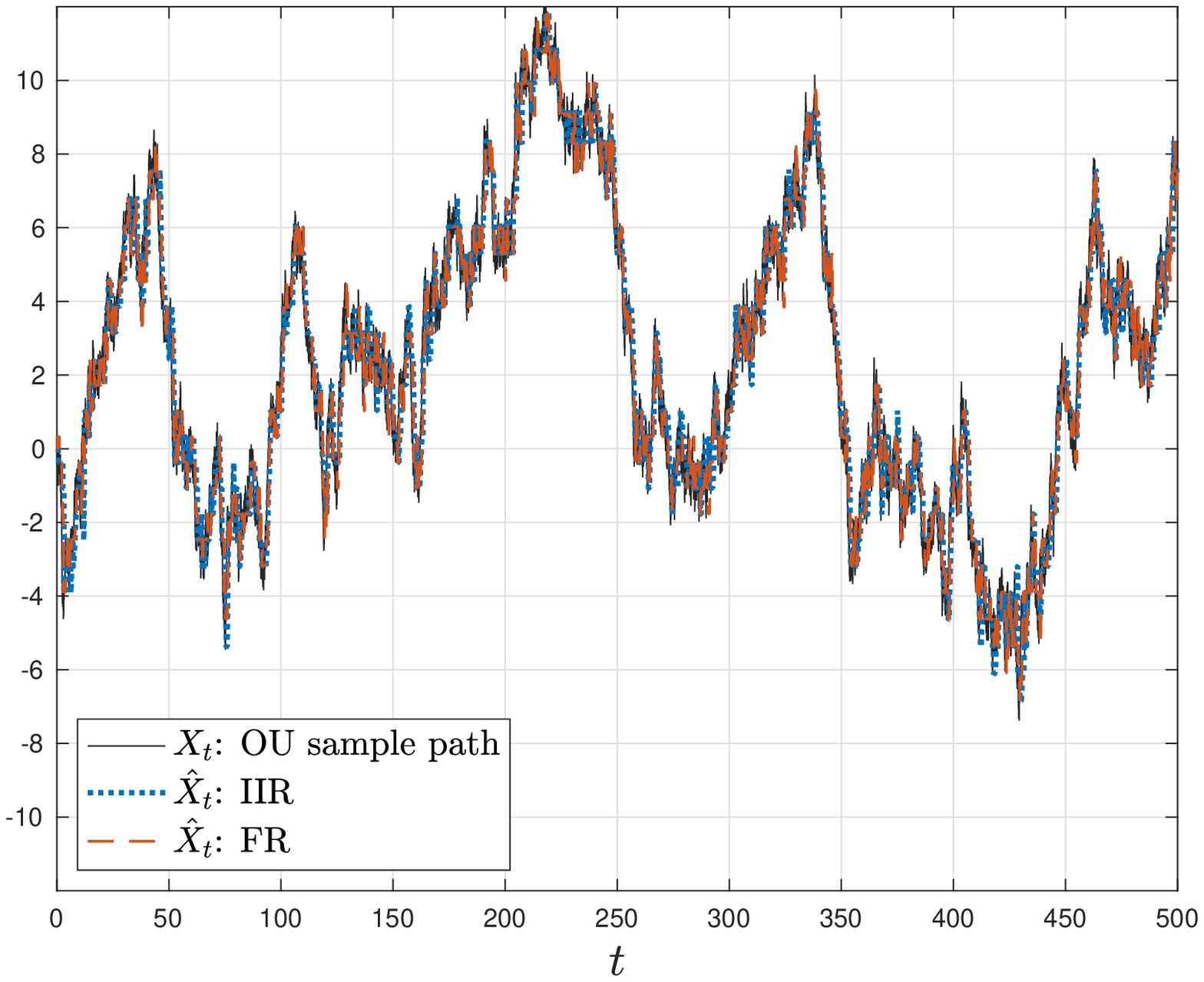}
\caption{Full view}
\label{fig_tracking_beta_15_theta_01_eps_1_FULL}
\end{subfigure}
\begin{subfigure}{.5\textwidth}
\center
\includegraphics[scale=.4]{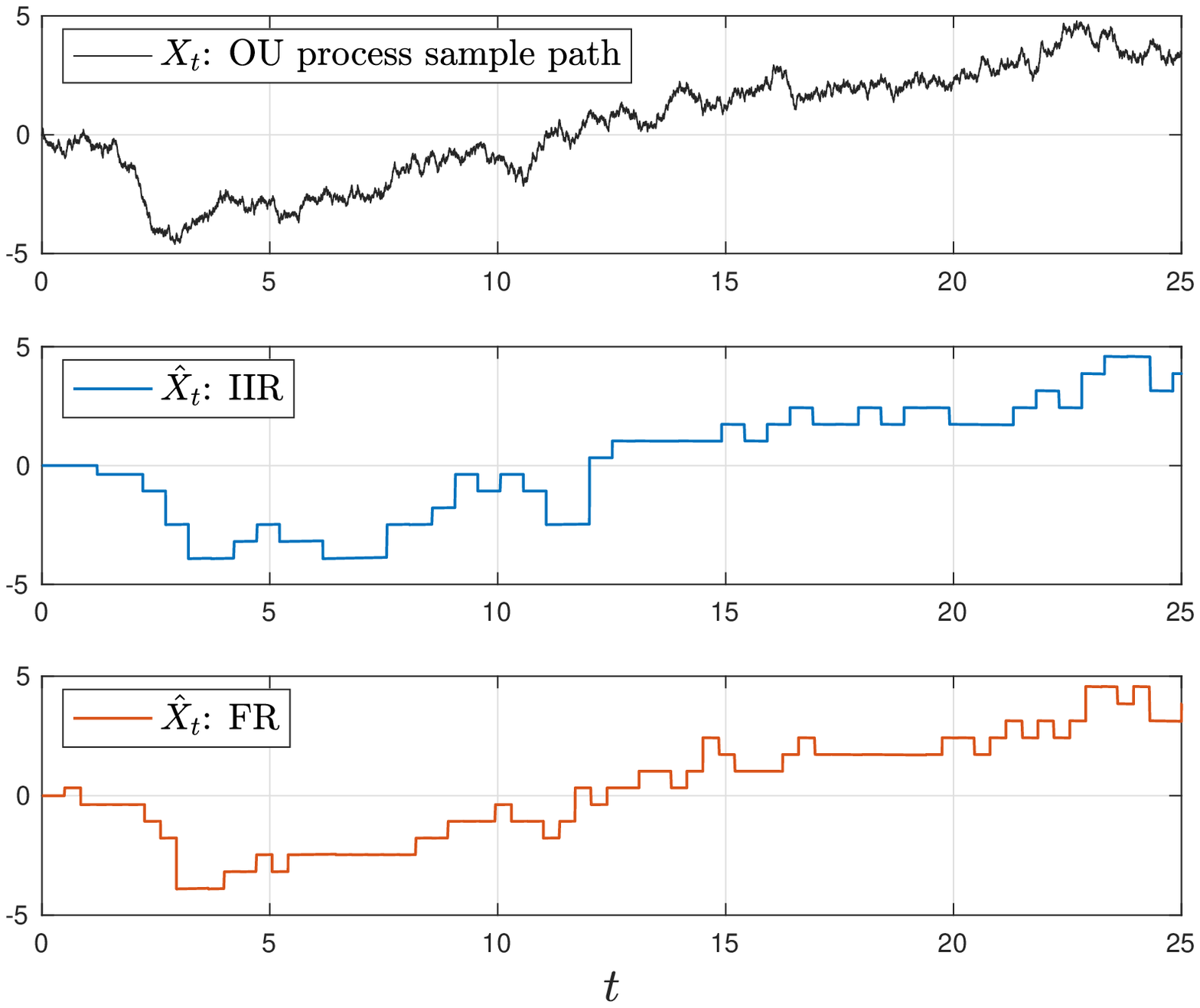}
\caption{Zoomed view}
\label{fig_tracking_beta_15_theta_01_eps_1_ZOOM}
\end{subfigure}
\caption{Tracking an OU sample path by generating an MMSE estimate using IIR and FR. We fix $\beta=0.15$, $\theta=0.01$ and $\epsilon=0.1$, and use (the optimal) $\ell=5$ and $n=7$.}
\label{fig_tracking_beta_15_theta_01_eps_1}
\end{figure}

The results are shown in Fig.~\ref{fig_tracking_beta_15_theta_01_eps_1}. The full view in Fig.~\ref{fig_tracking_beta_15_theta_01_eps_1_FULL} shows that both IIR and FR are able to allow the receiver to produce MMSE estimates that closely-track the original OU sample path. While the zoomed view in Fig.~\ref{fig_tracking_beta_15_theta_01_eps_1_ZOOM} shows the specifics of how the MMSE estimates look like. Empirically, the MSE for this sample path is $\approx0.87$ for IIR and $\approx0.74$ for FR, which are close to the theoretical values of the long-term average MMSE evaluated in Fig.~\ref{fig_iir_and_fr_vs_ell_theta_01_beta_15_eps_1_4}. This shows the ability of our techniques to achieve {\it timely tracking} of the process.

\section{Conclusions and Extensions}

A study of the effects of sampling, quantization and coding over noisy channels on MMSE estimates of an OU process has been presented. Focusing on MMSE quantizers, together with IIR and FR codes, a joint optimization problem of when to take new samples, how many quantization and codeword bits to use, has been formulated and solved. A fixed non-zero processing time has been considered at the receiver, modeling mainly decoding and feedback transmission times. It is shown how finely tuning the sampling and transmission times could make us of the processing time to send new data in order to save time in case decoding fails. Through numerical evaluations, it is shown that IIR performs relatively better than FR with small processing times, and vice versa, and so neither coding scheme dominates. It is also shown that the techniques developed in this paper can achieve timely tracking of the original process at the receiver's end.

In this work, the focus has been on signal-independent sampling policies. As an extension, one could develop techniques that work for {\it signal-dependent} sampling policies instead, in which the state of the OU process is observable to the sampler. While this is expected to produce better results, this comes with the challenge of {\it jointly} designing an MMSE quantizer {\it and} deriving an MMSE estimate at the receiver in this case. More generally though, there has been a separation-based quantization and coding methodology followed in this work, with focusing on two relatively-simple coding strategies. One could investigate the benefits of jointly optimizing the quantizer and the transmission code being used to convey the samples to the receiver with the smallest MMSE, which can be done for either signal-independent or signal-dependent sampling policies. Some structural properties of the tracked process may also guide the joint design in this case, as in, e.g., the sparse signal framework of \cite{cohen2019sparseQuant}. Finally, one can also extend the notion of fixed processing times to more practical models that take into consideration the code rate being used, together with noise in the feedback channel. As a more direct extension focusing on this point, one may consider {\it random} processing times, which calls for the investigation of whether it is useful to generate a new sample while an old one is still being processed if the processing time becomes relatively large.

\appendix

\subsection{Proof of Theorem~\ref{thm_iir_main_result}} \label{apndx_pf_iir_main_result}

We introduce the following Lagrangian \cite{boyd}:\footnote{Using the monotonicity of $g(\cdot)$, it can be shown that problem (\ref{opt_iir_aux}) is convex.}
\begin{align}
\mathcal{L}=\mathbb{E}\left[\int_{\overline{D}}^{\overline{D}+w\left(\overline{Y}\right)+Y}g\left(t-\overline{S}\right)dt\right]-\lambda\mathbb{E}\left[w\left(\overline{Y}\right)+Y\right]-\sum_{\bar{y}}w(\bar{y})\eta(\bar{y}),
\end{align}
where $\eta(\bar{y})$ is a Lagrange multiplier. Using Leibniz rule, we take the functional derivative with respect to $w(\bar{y})$ and equate to $0$ to get
\begin{align}
\mathbb{E}\left[g\left(\bar{y}+w^*(\bar{y})+Y\right)\right]=\lambda+\frac{\eta(\bar{y})}{\mathbb{P}\left(\overline{Y}=\bar{y}\right)}.
\end{align}
Since $g$ is increasing, the left hand side above is therefore an increasing function of $w^*(\bar{y})$, which we denote $G_{\bar{y}}\left(\cdot\right)$ in the theorem statement. Now, if $\lambda\leq G_{\bar{y}}(0)$, then we must have $\eta(\bar{y})>0$, and hence $w^*(\bar{y})=0$ by complementary slackness \cite{boyd}. Conversely, if $\lambda> G_{\bar{y}}(0)$, then we must have $w^*(\bar{y})>0$, and hence $\eta(\bar{y})=0$ also by complementary slackness. In the latter case, $w^*(\bar{y})=G_{\bar{y}}^{-1}(\lambda)$. Finally, observe that $\lambda\leq G_{\bar{y}}(0)\iff G_{\bar{y}}^{-1}(\lambda)\leq0$. This concludes the proof.

\subsection{Proof of Theorem~\ref{thm_fr_main_result}} \label{apndx_pf_fr_main_result}

We first simplify the terms of the objective function of (\ref{opt_fr_aux}). Using iterated expectations, it can be shown that
\begin{align}
\mathbb{E}\left[\sum_{j=1}^Mw_j+M\bar{n}\right]=\sum_{j=1}^\infty w_j(1-p_0)^{j-1}+\frac{\bar{n}}{p_0}.
\end{align}
Now let us define
\begin{align}
\zeta_m\left({\bm w}_1^m\right)\triangleq\int_{\overline{D}}^{\overline{D}+\sum_{j=1}^mw_j+m\bar{n}}g\left(t-\overline{S}_{\bar{M}}\right)dt
\end{align}
and, leveraging iterated expectations on the first term of (\ref{opt_fr_aux}), introduce the following Lagrangian:\footnote{Again, as mentioned above, it can be shown that problem (\ref{opt_fr_aux}) is convex using monotonicity of $g(\cdot)$.}
\begin{align}
\mathcal{L}=\sum_{m=1}^\infty \zeta_m\left({\bm w}_1^m\right)(1-p_0)^{m-1}p_0-\lambda\sum_{j=1}^\infty w_j(1-p_0)^{j-1}-\lambda\frac{\bar{n}}{p_0}-\sum_{j=1}^\infty w_j\eta_j,
\end{align}
where $\eta_j$'s are Lagrange multipliers. Now observe that, using Leibniz rule, it holds for $j\leq m$ that
\begin{align}
\frac{\partial \zeta_m\left({\bm w}_1^m\right)}{\partial w_j}=g\left(\bar{n}+\sum_{j=1}^mw_j+m\bar{n}\right).
\end{align}
Taking derivative of the Lagrangian with respect to $w_j$ and equating to $0$, we use the above to get
\begin{align} \label{eq_fr_pf_wj}
\sum_{m=j}^\infty g\!\left(\!\bar{n}+\sum_{j=1}^mw_j+m\bar{n}\!\right)\!(1-p_0)^{m-j}p_0\!=\lambda\!+\!\frac{\eta_j}{(1\!-\!p_0)^{j-1}}.
\end{align}
Next, let us substitute $j=k$ and $j=k+1$ above, $k\geq1$, subtract them from each other, and rearrange to get
\begin{align}
g\left(\bar{n}+\sum_{j=1}^kw_j+k\bar{n}\right)=\lambda+\frac{\eta_k-\eta_{k+1}}{(1-p_0)^{k-1}p_0}.
\end{align}
Since $g(\cdot)$ is increasing, and $\lambda$ is fixed, $\left\{\frac{\eta_k-\eta_{k+1}}{(1-p_0)^{k-1}p_0}\right\}$ is increasing. From there, one can conclude that $\eta_j>0,~j\geq2$ must hold. Hence, by complementary slackness, $w_j^*=0,~j\geq2$ \cite{boyd}. Using (\ref{eq_fr_pf_wj}) for $j=1$, the optimal $w_1^*$ now solves
\begin{align}
G\left(w_1^*\right)=\lambda+\eta_1,
\end{align}
where $G(\cdot)$ is as defined in the theorem statement. Observe that $G(\cdot)$ is increasing and therefore the above has a unique solution. Proceeding similarly as in the proof of Theorem~\ref{thm_iir_main_result}, if $\lambda\leq G(0)$, then we must have $\eta_1>0$, and hence $w_1^*=0$ by complementary slackness; conversely, if $\lambda>G(0)$, then we must have $w_1^*>0$, and hence $\eta_1=0$ by complementary slackness as well \cite{boyd}. In the latter case, $w_1^*=G^{-1}(\lambda)$. Finally, observe that $\lambda\leq G(0)\iff G^{-1}(\lambda)\leq0$. This concludes the proof of the first part of the theorem.

To show the second part, all we need to prove now is that $G^{-1}\left(\lambda^*_{FR}\right)\leq0$, or equivalently that $\lambda^*_{FR}\leq G(0)$. Toward that end, observe that $p_{FR}(\lambda)$ is decreasing, and therefore if $p_{FR}\left(G(0)\right)\leq0$ then the premise follows. Now for $\lambda=G(0)$ we know from the first part of the proof that $w_1^*=0$. Thus,
\begin{align}
p_{FR}\left(G(0)\right)=&\sum_{m=1}^\infty\zeta_m\left(0\right)(1-p_0)^{m-1}p_0-G(0)\frac{\bar{n}}{p_0} \\
=&\mathbb{E}\left[\int_{\overline{D}}^{\overline{D}+M\bar{n}}g\left(t-\overline{S}_{\bar{M}}\right)dt\right]-G(0)\mathbb{E}\left[M\right]\bar{n} \\
=&\mathbb{E}\left[\int_{0}^{M\bar{n}}g\left(\bar{n}+t\right)dt\right]-\mathbb{E}\left[\int_0^{M\bar{n}}G(0)dt\right] \label{eq_fr_pf_w1_0_1} \\
=&\mathbb{E}\left[\int_{0}^{M\bar{n}}\mathbb{E}\left[g\left(\bar{n}+t\right)-g\left(\bar{n}+M\bar{n}\right)\right]dt\right],\label{eq_fr_pf_w1_0_2}
\end{align}	
where (\ref{eq_fr_pf_w1_0_1}) follows by change of variables and (\ref{eq_fr_pf_w1_0_2}) follows by definition of $G(\cdot)$. Finally, observe that by monotonicity of $g(\cdot)$, (\ref{eq_fr_pf_w1_0_2}) is non-positive. This concludes the proof.

\subsection{Proof of Lemma~\ref{thm_iir_enhanced}} \label{apndx_pf_iir_enhanced}

Let us consider the $i$th epoch. We prove the lemma by computing the channel delay experienced by the enhanced scheme for some realization of $r_i$. The proof can be better-conveyed graphically through Figs.~\ref{fig_iir_enhanced_beta} and~\ref{fig_iir_enhanced_Tb} below. We will consider two cases as follows.

\subsubsection{$\beta\leq T_b$}

In this case, the first feedback following the initial $nT_b$ time units is received while the first IR bit is still being transmitted. If it is an ACK, then the transmitter stops and cuts off the current IR bit transmission and ends the epoch with a channel delay of $nT_b+\beta$. Otherwise, if it is a NACK, then the receiver will begin re-processing with a codeword of length $n+1$ after exactly $T_b-\beta$ time units from the time the feedback is received. Simultaneously, the transmitter will send the second IR bit. The process is repeated till an ACK is received.

\begin{figure}[t]
\center
\includegraphics[scale=.5]{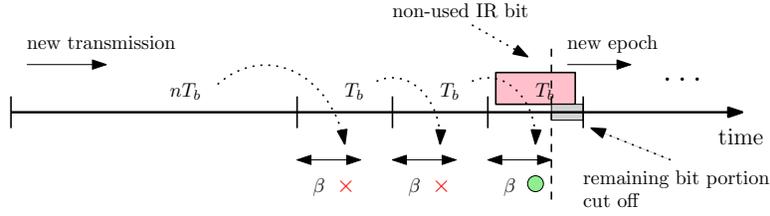}
\caption{Example sample path during the $i$th epoch using the enhanced IIR scheme when $\beta\leq T_b$. In this example $r_i=2$, and so the third IR bit is non-used and its remaining portion is cut off to start a new epoch. Red crosses denote failed decoding attempts and the green circle denotes success.}
\label{fig_iir_enhanced_beta}
\end{figure}

In general, an ACK will be received after $r_i$ IR bits, and the $(r_i+1)$th bit will be cut off (this bit will be a non-used IR bit). This ends the epoch with a channel delay of exactly
\begin{align}
nT_b+r_i\beta+r_i(T_b-\beta)+\beta=\bar{n}+r_iT_b,
\end{align}
which saves $r_i\beta$ time units compared to the original IIR scheme that waits for feedback before sending IR bits. An example sample path is shown in Fig.~\ref{fig_iir_enhanced_beta}.

\subsubsection{$\beta>T_b$}

Different from the $\beta\leq T_b$ case, the transmitter can now possibly fit more than one IR bit while the receiver is processing previously-received bits. Specifically, a total of $\lfloor\beta/T_b\rfloor$ IR bits would be received by the end of the first decoding attempt, a total of $\lfloor2\beta/T_b\rfloor$ IR bits would be received by the end of the second decoding attempt, and so on.

\begin{figure}[t]
\center
\includegraphics[scale=.5]{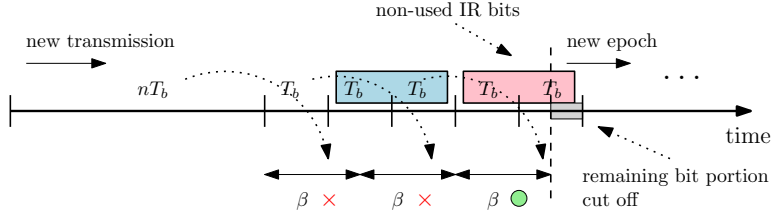}
\caption{Example sample path during the $i$th epoch using the enhanced IIR scheme when $\beta>T_b$. In this example $r_i=2$ and $\beta=1.5T_b$, and so the final two IR bits are non-used and the remaining bit portion is cut off to start a new epoch. Red crosses denote failed decoding attempts and the green circle denotes success.}
\label{fig_iir_enhanced_Tb}
\end{figure}

Now let $\kappa_i$ be as defined in the lemma. This way, the required IR bits for successful decoding will be available after exactly $\kappa_i\beta$ time units following the initial $nT_b$ time units, and an ACK will be fed back $\beta$ time units afterwards. By the time an ACK is received, there would be already some extra IR bits sent to the receiver that were not needed in decoding (these will be non-used IR bits). In addition, there could be an extra bit portion that needs to be cut off belonging to an IR bit that is being transmitted while the ACK is received; this occurs if $(\kappa_i+1)\beta>\lfloor(\kappa_i+1)\beta/T_b\rfloor T_b$. This ends the epoch with a channel delay of exactly
\begin{align}
nT_b+\kappa_i\beta+\beta=\bar{n}+\kappa_i\beta
\end{align}
which saves $r_iT_b+(r_i-\kappa_i)\beta$ time units. An example sample path is shown in Fig.~\ref{fig_iir_enhanced_Tb}.

\subsection{Proof of Lemma~\ref{thm_fr_enhanced}} \label{apndx_pf_fr_enhanced}

Let $L$ denote the epoch length, and let $Q$ denote the cumulative age-penalty in the epoch given by
\begin{align}
Q=\int_{\overline{D}}^{\overline{D}+L}g\left(t-\overline{S}_{\bar{M}}\right)dt.
\end{align}
Recalling the definition of $\delta$, our goal is to characterize $\mathbb{E}[L]$ and $\mathbb{E}[Q]$ in terms of $\delta$ and solve the following optimization problem to find $\delta^*$:
\begin{align} \label{opt_fr_enhanced}
\min_{0\leq\delta\leq\beta}\quad\frac{\mathbb{E}[Q]}{\mathbb{E}[L]}.
\end{align}
Similar to the proof of Lemma~\ref{thm_iir_enhanced} in Appendix~\ref{apndx_pf_iir_enhanced}, our proof methodology is made clearer through Figs.~\ref{fig_fr_enhanced_beta} and~\ref{fig_fr_enhanced_Tb}, and we will consider two cases as follows.

\subsubsection{$\beta\leq nT_b$}

\begin{figure}
\center
\includegraphics[scale=.5]{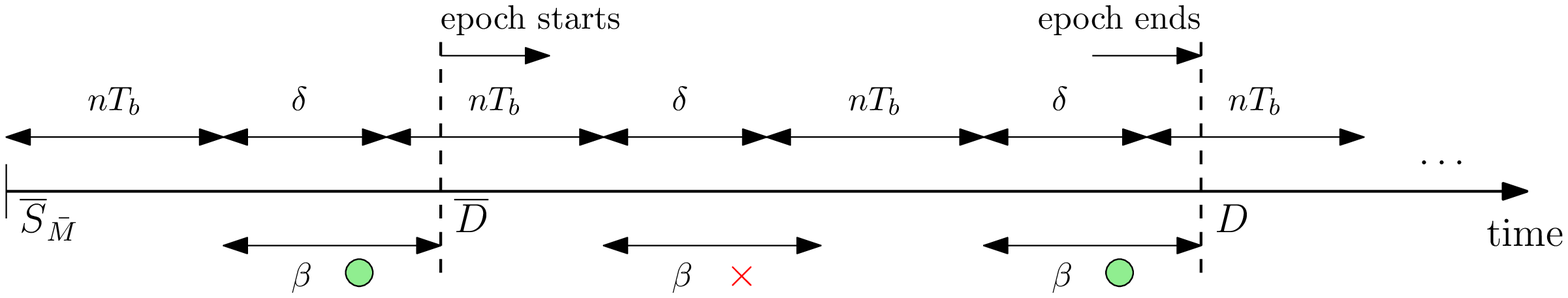}
\caption{Example sample path during an epoch using the enhanced FR scheme when $\beta\leq nT_b$. In this example $M=2$, and so it takes two transmissions to succeed. The red cross denotes a failed decoding attempt and green circles denote success.}
\label{fig_fr_enhanced_beta}
\end{figure}

In this case, we need to show $\delta^*=0$. Right before the epoch starts, there would be $\left\lfloor(\beta-\delta)/T_b\right\rfloor$ bits (belonging to a new message) already available. The first decoding attempt in the epoch, therefore, occurs after $nT_b-\beta+\delta$ time units from the epoch's start time. If this decoding attempt is successful, an ACK will be fed back after $\beta$ time units. Otherwise, a new message will be transmitted through the same manner again, see Fig.~\ref{fig_fr_enhanced_beta}. From the figure, one can see that the epoch length is given by
\begin{align}
L=&\left(\left(nT_b-\beta+\delta\right)+\beta\right)M \\
=&\left(nT_b+\delta\right)M,
\end{align}
and therefore
\begin{align}
\mathbb{E}[L]=&\frac{nT_b+\delta}{p_0}, \label{eq_exp_L_fr_enhanced_beta} \\
\mathbb{E}[Q]=&\sum_{m=1}^\infty\left(\int_{\overline{D}}^{\overline{D}+\left(nT_b+\delta\right)m}g\left(t-\overline{S}_{\bar{M}}\right)dt\right)(1-p_0)^{m-1}p_0. \label{eq_exp_Q_fr_enhanced_beta}
\end{align}

Next, we follow Dinkelbach's approach \cite{dinkelbach-fractional-prog} to solve problem (\ref{opt_fr_enhanced}) and introduce the auxiliary problem
\begin{align}
q(\lambda)\triangleq\min_{0\leq\delta\leq\beta}\quad\mathbb{E}[Q]-\lambda\mathbb{E}[L]
\end{align}
for some $\lambda\geq0$. We introduce the following Lagrangian for such problem \cite{boyd}:
\begin{align}
\mathcal{L}=&\mathbb{E}[Q]-\lambda\mathbb{E}[L]-\eta\delta+\omega(\delta-\beta),
\end{align}
where $\eta$ and $\omega$ are Lagrange multipliers. Now using (\ref{eq_exp_L_fr_enhanced_beta}) and (\ref{eq_exp_Q_fr_enhanced_beta}), we take the derivative with respect to $\delta$ to get
\begin{align}
\frac{d\mathcal{L}}{d\delta}=&\sum_{m=1}^\infty mg\left(\overline{D}+\left(nT_b+\delta\right)m-\overline{S}_{\bar{M}}\right)(1-p_0)^{m-1}p_0-\frac{\lambda}{p_0}-\eta+\omega \\
=&\sum_{m=1}^\infty mg\left(\bar{n}+\left(nT_b+\delta\right)m\right)(1-p_0)^{m-1}p_0-\frac{\lambda}{p_0}-\eta+\omega \\
\triangleq&H(\delta)-\frac{\lambda}{p_0}-\eta+\omega.
\end{align} 
Therefore, the optimal $\delta^*$ solves
\begin{align}
H\left(\delta^*\right)=\frac{\lambda}{p_0}+\eta-\omega.
\end{align}
Note that $H(\delta)$ is increasing in $\delta$ by monotonicity of $g(\cdot)$. Hence, if $\lambda<p_0H(0)$ then we must have $\eta>0$, which implies by complementary slackness that $\delta^*=0$.

We now proceed similarly as in the second part of the proof of Theorem~\ref{thm_fr_main_result} in Appendix~\ref{apndx_pf_fr_main_result}. Specifically, since the optimal $\lambda^*$ satisfies $q(\lambda^*)=0$ and $q(\lambda)$ is decreasing \cite{dinkelbach-fractional-prog}, it suffices to show that $q\left(p_0H(0)\right)<0$. Towards that end, we have
\begin{align}
q\left(p_0H(0)\right)=&\sum_{m=1}^\infty\left(\int_{\overline{D}}^{\overline{D}+nT_bm}g\left(t-\overline{S}_{\bar{M}}\right)dt\right)(1-p_0)^{m-1}p_0-p_0H(0)\frac{nT_b}{p_0} \\
<&\sum_{m=1}^\infty nT_bmg\left(\overline{D}+nT_bm-\overline{S}_{\bar{M}}\right)(1-p_0)^{m-1}p_0-H(0)nT_b \\
=&0,
\end{align}
where the inequality follows by monotonicity of $g(\cdot)$, and the last equality follows by definition of $H(\cdot)$.

\subsubsection{$\beta>nT_b$}

In this case, we need to show $\delta^*=\beta-nT_b$. We first argue that $\delta^*$ cannot be smaller than $\beta-nT_b$. To see this, observe that if $\delta^*<\beta-nT_b$, then there would be a codeword waiting in the receiver's queue for $\beta-nT_b-\delta^*$ time units after being completely received before it gets processed. One can strictly decrease the age-penalty in this case by acquiring {\it fresher} sample instead of the current one via pushing the sampling time exactly $\beta-nT_b-\delta^*$ time units forward and avoid the unnecessary idle waiting at the receiver. Thus, our goal now is to solve problem (\ref{opt_fr_enhanced}) over the new bound $\delta\in\left[\beta-nT_b,\beta\right]$.

\begin{figure}
\center
\includegraphics[scale=.75]{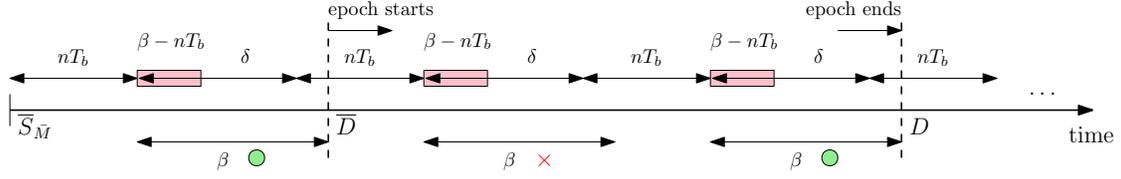}
\caption{Example sample path during an epoch using the enhanced FR scheme when $\beta>nT_b$. In this example $M=2$, and so it takes two transmissions to succeed. Light-red boxes represent the lower bound on $\delta$ (idle times). The red cross denotes a failed decoding attempt and green circles denote success.}
\label{fig_fr_enhanced_Tb}
\end{figure}

As in the previous case, and now that $\delta\geq\beta-nT_b$, there would also be $\left\lfloor(\beta-\delta)/T_b\right\rfloor$ bits available from a new message right before the epoch starts, and the first decoding attempt in the epoch would occur after $nT_b-\beta+\delta$ time units from the epoch's start time. This repeats until an ACK is fed back, see Fig.~\ref{fig_fr_enhanced_Tb}.

This gives rise to the exact same $\mathbb{E}[L]$ and $\mathbb{E}[Q]$ expressions in (\ref{eq_exp_L_fr_enhanced_beta}) and (\ref{eq_exp_Q_fr_enhanced_beta}), respectively. One can thus follow the same analysis for the $\beta\leq nT_b$ case to solve the optimization problem and reach the conclusion that $\delta^*$ should be equal to its lower bound, $\beta-nT_b$ in this case.

\subsection{Deriving Equations (\ref{eq_iir_enhanced_wait}) and (\ref{eq_fr_enhanced_mse})} \label{apndx_fr_enhanced_mse}

We derive the optimal waiting policy in (\ref{eq_iir_enhanced_wait}) by solving $G_{\bar{y}}\left(w^*(\bar{y})\right)=\lambda^*_{IIR}$ with $G_{\bar{y}}(\cdot)$ as defined in Theorem~\ref{thm_iir_main_result}, with $g(\cdot)\equiv h_\ell(\cdot)$, after replacing the random variable $Y$ with $\tilde{Y}$. That is,
\begin{align}
G_{\bar{y}}\left(w^*(\bar{y})\right)=&\mathbb{E}\left[h_{\ell}\left(\bar{y}+w^*(\bar{y})+\tilde{Y}\right)\right] \nonumber \\
=&\frac{\sigma^2}{2\theta}\left(1-\left(1-2^{-2\ell}\right)e^{-2\theta\left(\bar{y}+w^*(\bar{y})\right)}\mathbb{E}\left[e^{-2\theta\tilde{Y}}\right]\right) \nonumber \\
=&\lambda^*_{IIR},
\end{align}
whence (\ref{eq_iir_enhanced_wait}) directly follows by solving for $w^*(\bar{y})$ above and taking the non-negative part.

Next, we derive the long-term average MMSE expression in (\ref{eq_fr_enhanced_mse}) through basically evaluating the optimal $\mathbb{E}[L]$ and $\mathbb{E}[Q]$ in (\ref{eq_exp_L_fr_enhanced_beta}) and (\ref{eq_exp_Q_fr_enhanced_beta}), respectively, with $g(\cdot)\equiv h_\ell(\cdot)$, after substituting $\delta^*=\left[\beta-nT_b\right]^+$. First, we have
\begin{align}
\mathbb{E}[L]=&\frac{nT_b+\left[\beta-nT_b\right]^+}{p_0} \nonumber \\
=&\frac{K_{n,\beta}}{p_0}.
\end{align}
Next, we have
\begin{align}
\mathbb{E}[Q]=&\sum_{m=1}^\infty\left(\int_{\overline{D}}^{\overline{D}+\left(nT_b+\left[\beta-nT_b\right]^+\right)m}h_{\ell}\left(t-\overline{S}_{\bar{M}}\right)dt\right)(1-p_0)^{m-1}p_0 \nonumber \\
=&\sum_{m=1}^\infty\left(\int_{\overline{D}}^{\overline{D}+K_{n,\beta}m}\frac{\sigma^2}{2\theta}\left(1-\left(1-2^{-2\ell}\right)e^{-2\theta\left(t-\overline{S}_{\bar{M}}\right)}\right)dt\right)(1-p_0)^{m-1}p_0 \nonumber \\
=&\frac{\sigma^2}{2\theta}\left(\frac{K_{n,\beta}}{p_0}-\frac{\left(1-2^{-2\ell}\right)e^{-2\theta\bar{n}}}{2\theta}\left(1-\frac{p_0e^{-2\theta K_{n,\beta}}}{1-(1-p_0)e^{-2\theta K_{n,\beta}}}\right)\right) \nonumber \\
=&\frac{\sigma^2}{2\theta}\left(\frac{K_{n,\beta}}{p_0}-\frac{\left(1-2^{-2\ell}\right)e^{-2\theta\bar{n}}}{2\theta}\frac{1-e^{-2\theta K_{n,\beta}}}{1-(1-p_0)e^{-2\theta K_{n,\beta}}}\right).
\end{align}
Equation (\ref{eq_fr_enhanced_mse}) now directly follows via dividing $\mathbb{E}[Q]$ above by $\mathbb{E}[L]$.


\end{document}